\documentclass[acmsmall]{acmart}

\usepackage{tabularx}
\usepackage{multirow}
\usepackage{float}
\usepackage{subfig}
\usepackage{tikz}
\usepackage[normalem]{ulem}

\newcommand{\minitab}[2][l]{\begin{tabular}{#1}#2\end{tabular}}
\newcommand{\subhead}[1]{\vspace{2pt} \noindent {\bf #1}}
\newcommand{\secspace}{\vspace{0pt}}
\newcommand{\subsecspace}{\vspace{0pt}}
\newcommand{\etal}{\emph{et al.\ }}

\newcommand{\simnet}{\textsc{SimNet}}

\AtBeginDocument{%
  \providecommand\BibTeX{{%
    \normalfont B\kern-0.5em{\scshape i\kern-0.25em b}\kern-0.8em\TeX}}}

\acmJournal{POMACS}
\acmVolume{1}
\acmNumber{1}
\acmArticle{1}
\acmMonth{6}

\begin{document}

\title{SimNet: Accurate and High-Performance Computer Architecture Simulation using Deep Learning}

\author{Lingda Li}
\email{lli@bnl.gov}
\affiliation{%
  \institution{Brookhaven National Laboratory}
  \city{Upton}
  \state{NY}
  \country{USA}
}

\author{Santosh Pandey}
\email{spande1@stevens.edu}
\affiliation{%
  \institution{Stevens Institute of Technology}
  \city{Hoboken}
  \state{NJ}
  \country{USA}
}

\author{Thomas Flynn}
\email{tflynn@bnl.gov}
\affiliation{%
  \institution{Brookhaven National Laboratory}
  \city{Upton}
  \state{NY}
  \country{USA}
}

\author{Hang Liu}
\email{hliu77@stevens.edu}
\affiliation{%
  \institution{Stevens Institute of Technology}
  \city{Hoboken}
  \state{NJ}
  \country{USA}
}

\author{Noel Wheeler}
\email{nwheeler@lps.umd.edu}
\affiliation{%
  \institution{Laboratory for Physical Sciences}
  \city{College Park}
  \state{MD}
  \country{USA}
}

\author{Adolfy Hoisie}
\email{ahoisie@bnl.gov}
\affiliation{%
  \institution{Brookhaven National Laboratory}
  \city{Upton}
  \state{NY}
  \country{USA}
}


\begin{abstract}
While cycle-accurate simulators are essential tools for architecture research,
design, and development, their practicality is limited by an extremely long
time-to-solution for realistic applications under investigation.
This work describes a concerted effort, where machine learning (ML) is used to
accelerate microarchitecture simulation.
First, an ML-based instruction latency prediction framework that accounts for
both static instruction properties and dynamic processor states is constructed.
Then, a GPU-accelerated parallel simulator is implemented based on the proposed
instruction latency predictor, and its simulation accuracy and throughput are
validated and evaluated against a state-of-the-art simulator.
Leveraging modern GPUs, the ML-based simulator outperforms traditional
CPU-based simulators significantly.
\end{abstract}



\maketitle

\secspace
\section{Introduction}
\label{sect:intro}
\secspace

Adopted extensively in computer architecture research and engineering,
cycle-accurate discrete-event simulators (DES) enable new architectural ideas,
as well as design space exploration.
DES is composed of distinct modules that mimic the behavior of different
hardware components.
On certain events (e.g., advancing a cycle), these individual components and
their interactions are simulated to imitate the behavior of processors.
Unfortunately, DES is extremely computationally demanding, markedly diminishing
its practicality and applicability at full system and  application scales.
Typical simulations using the state-of-the-art gem5 simulator \cite{gem5}
execute at speeds of hundreds of kilo instructions per second on modern CPUs,
about four to five orders of magnitude slower than native execution.
In this context, it would require weeks or months to simulate a realistic
application that only takes a couple minutes to execute on real hardware.
To expand the practical limits of traditional simulation, design space
exploration necessitates a multitude of simulations across various applications
and design parameters under consideration.


Many efforts have been made to improve traditional simulation speed through
software engineering optimizations \cite{sandberg2015, cmpsim, marss},
multi/many-core simulation parallelization \cite{sst, ZSim}, and statistical
approaches \cite{ASPLOS02:SimPoint, SIGMETRICS03:SimPoint, ISCA03:SMARTS}.
Among them, parallelization is a promising direction due to the broad
availability of massive parallel accelerators such as GPUs nowadays.
However, one fundamental limitation of these traditional simulators is they are
intrinsically difficult to parallelize because of the heterogeneous nature of
distinct components, frequent interactions between components, and irregular
behaviors.
As a result, existing parallel simulators focus on a coarse parallelization
strategy, where the simulation of individual cores is parallelized \cite{sst,
ZSim}.
These simulators have limited scalability and cannot leverage modern parallel
accelerators.

In the meantime, machine learning (ML) advances have led to remarkable
achievements in many domains, and using ML for analytical performance modeling
is significant and growing.
Considerable research has been done to predict application performance
\cite{ASPLOS06:Ipek, HPCA07:Lee, PPoPP07:Lee, agbarya20-mosalloc,
MICRO15:Ardalani, SBAC-PAD14:Baldini, TECS17:Oneal, DAC16:Zheng}.
The major limitation of these application-centric methods is they are
program/input dependent, which means ML models need to be trained for
individual program and input combinations.
As a result, their flexibility is limited compared to simulation-based
approaches.

This paper aims to explore the possibility of an ML-based computer architecture
simulation approach given the following reasons.
First, ML models, especially deep neural networks, have been proved to be
excellent function approximators in many domains, from computer vision to
scientific computing \cite{transformer, ICML19:EfficientNet, alphafold,
SC20:Jia}.
We expect they can also be applied to approximate the complex and implicit
latency calculations that are essential to computer architecture simulation.
Second, ML-based simulation is more flexible compared with ML-based analytical
modeling because it does not require training per program/input.
Moreover, an ML-based simulator could bring performance advantages because ML
inferences are highly parallel, and state-of-the-art accelerators (e.g., GPU;
TPU \cite{tpu}) and software infrastructures \cite{pytorch, tensorflow,
tensorrt} are well optimized for such tasks.

Motivated by these potentials, we establish the first ML-based architecture
simulator, {\em \simnet{}}.
\simnet{} is a novel instruction-centric simulation framework that decomposes
program simulation into individual instruction latency and uses tailored ML
techniques for instruction latency prediction.
Program performance is obtained by combining the latency prediction results of
all executed instructions.
\simnet{} achieves noticeably higher simulation throughput while maintaining
the same level of simulation accuracy because: 1) it abstracts the simulated
processor as a whole and eliminates the need to simulate individual components
within the processor, and 2) it is well optimized to execute on GPUs
efficiently.
Moreover, \simnet{} can simulate complex processor architectures and realistic
application workloads with billions/trillions of instructions.
The source code of \simnet{} is available at
\url{https://github.com/lingda-li/simnet}.

This work's contributions to the science and practice of simulation include:
\begin{itemize}
\item We propose an ML framework to predict instruction latency accurately
(Section \ref{sect:ml}).
The proposed framework accounts for both static instruction properties and
dynamic processor behaviors, and extensive ML models are evaluated to balance
between the prediction accuracy and speed.
\item We propose an instruction-centric architecture simulator that is built
upon ML-based instruction latency predictors (Section \ref{sect:sim}).
Evaluated using a realistic benchmark suite and on full microprocessor
architectures, we demonstrate that the proposed approach simulates programs
faithfully compared with the discrete-event simulator it learns from.
To the best of our knowledge, the proposed framework is the first of its kind
and could set the stage for developing alternative tools for architecture
researchers and engineers.
\item We also prototype a GPU-accelerated ML-based simulator (Section
\ref{sect:sim:accel}).
It improves the simulation throughput up to $76\times$ compared to its
traditional counterpart.
It also achieves a higher throughput per watt thanks to GPU's power efficiency
advantage.
\end{itemize}

\begin{figure*}[t]
\centering
\includegraphics[width=0.8\linewidth]{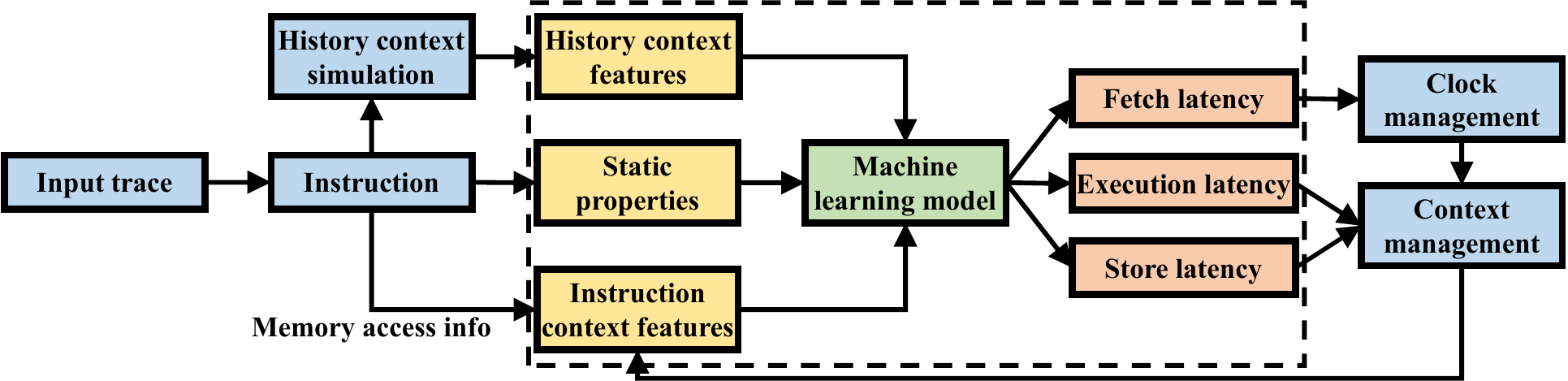}
\caption{ML-based simulation workflow. The ML-based instruction latency
predictor is shown in green, and its input and output are in yellow and orange,
respectively (Section \ref{sect:ml}). Other simulator components are in blue
(Section \ref{sect:sim}).}
\label{fig:mlsim}
\end{figure*}

\subhead{Related Work.}
Ithemal \cite{mendis2019ithemal} represents the closest related work to this
effort, proposing a long short-term memory (LSTM) model to predict the
execution latency of static basic blocks.
However, Ithemal has three major limitations: 1) it targets a simplified
processor model without branch prediction and cache/memory hierarchies; 2) it
can only predict the performance of basic blocks with a handful of
instructions, while real-world simulation executes billions or trillions of
instructions; and 3) it simulates instructions at a pace of thousands of
instructions per second, which is significantly slower than traditional
simulators.
As a result, unlike \simnet{}, Ithemal can neither simulate real-world
processors nor applications, and it is infeasible for realistic computer
architecture simulation.
Section \ref{sect:ml:accuracy} will quantitatively compare Ithemal with
\simnet{}, and Section \ref{sect:relatedwork} will discuss other related works.

\subhead{Scope of Work.}
This paper focuses on simulating out-of-order superscalar CPUs, which employ
technologies such as multi-issue, out-of-order scheduling, and speculative
execution to exploit instruction-level parallelism.
We posit the proposed simulation methodology also is applicable to other
processor architectures, which usually are less challenging to simulate.
In addition, we constrain the scope to the prediction of program performance
and single-thread program simulation, leaving multiple-thread/program
simulation for future work.
Traditional simulators also produce additional metrics other than performance,
such as energy consumption.
While it is reasonable to assume the proposed method is applicable to such
prediction as well, these metrics are not considered in this work.



\secspace
\section{ML-based Latency Prediction}
\label{sect:ml}
\secspace

Figure \ref{fig:mlsim} shows the workflow of \simnet{}.
\simnet{} is built around an ML-based instruction latency prediction framework
(the dashed box in Figure \ref{fig:mlsim}), and this section will describe its
design.

\subsecspace
\subsection{Factors that Determine Instruction Latency}
\label{sect:mot:factors}
\subsecspace

Successful instruction latency prediction by an ML model is contingent upon
capturing all factors that impact latency in its design and implementation.
These factors can be summarized into three categories.

\subhead{Static Instruction Properties.}
These properties describe the basics of an instruction, including the operation
types, source/destination registers, etc.
They guide how an instruction is executed in a processor.
For instance, the type of instruction determines its computation resource
(e.g., function units; register files) and synchronization requirements (e.g., memory
barriers).




\subhead{Dynamic Processor States.}
Besides its static properties, the latency of an instruction largely depends on
the states of all processor components (e.g., register files; caches) at its
execution time.
We refer to these states as {\em contexts} and further distill them into two
categories.

\textit{Instruction Context}:
Many contexts relate to other concurrently running instructions in the
processor, referred to as {\em instruction context} in this paper.
For example, whether the desired execution unit is available depends on whether
there is a co-running instruction of the same type using it currently, and if a
source register can be read immediately depends on whether the previous
instruction that writes the same register has finished.
We argue that instruction context can be determined given all concurrently
running instructions, named {\em context instructions} in this paper.
The processor capacity decides the maximal number of context instructions.

\textit{History Context}:
The remaining hardware contexts depend on events that happened in the long-term
execution history.
Cache, translation lookaside buffer (TLB), and branch predictor states belong
to this category.
For example, whether a memory load hits in the L2 cache depends on when the
same cache line was last accessed, and branch prediction results hinge on
the branch execution history.
Traditional simulators employ lookup tables (e.g., cache tag array; branch
target predictor) to keep track of such states.
We refer to them as {\em history context}.

\subsecspace
\subsection{Framework Formulation}
\label{sect:mot:io}
\subsecspace

\begin{table*}[t]
\footnotesize
\centering
\begin{tabularx}{\linewidth}{|l|X|}
\hline
{\bf Impact Factor} & {\bf Features} \\
\hline
\hline
Static properties & 13 operation features (function type, direct/indirect branch, memory barrier, etc.); 14 register indices (8 sources and 6 destinations) \\
\hline
Instruction context & 27 static properties; 14 history context features; 1 residence latency; 1 execution latency; 1 store latency; 5 memory dependency flags to indicate if it shares the same instruction/data address/cache line/page with the current instruction \\
\hline
History context & 1 branch misprediction flag; 1 fetch level; 3 fetch table walking levels; 2 fetch caused writebacks; 1 data access level; 3 data access table walking levels; 3 data access caused writebacks \\
\hline
\end{tabularx}
\caption{Input features for various instruction latency impact factors.}
\label{tbl:features}
\end{table*}

With these impact factors, we are ready to build an instruction latency
prediction framework.
The framework aims to balance between two competing goals: to predict
instruction latency {\em accurately} and {\em swiftly}.
An ML-based instruction latency predictor (the green box in Figure
\ref{fig:mlsim}) is the center of the framework, which captures the
impact of input features.
Its inputs (yellow boxes) take into account the aforementioned impact factors.
Table \ref{tbl:features} summarizes the input features, which are divided into
three categories based on which impact factor they model as introduced below.

\subhead{Modeling Static Instruction Properties.}
The top row of Table \ref{tbl:features} lists the static instruction properties
used as the input features of the ML model, including 13 operation features and
14 source and destination register indices.
They are well known to computer architects and can be extracted from the
instruction encoding directly.

\subhead{Modeling Instruction Context.}
To account for the impact of concurrently running instructions, the key is to
model their {\em relationships} with the to-be-predicted instruction.
Such relationships include resource competition, register dependency, and
memory dependency.
We call these concurrently running instructions, {\em context instructions}.
Formally, the context instructions are those instructions present in the
processor when a particular instruction is about to be fetched.
In theory, instructions issued after the current instruction also can influence
its execution.
However, these cases are rare, and we only include previous instructions for
practicality.
The middle row of Table \ref{tbl:features} shows input features per context
instruction.

For resource competition and register dependency, it is sufficient to provide
the static properties of context instructions (i.e., their operation features
and register indices).
The ML model is responsible for deducing the resource competition using their
operation features and the register dependency by comparing register indices.

To model the memory dependency (including instruction fetch and data access),
one solution is to provide memory access addresses as parts of input features
and leave the rest work to the ML model.
Unlike register indices, memory access addresses spread across a much wider
range in a typical 64-bit address space.
Thus, having addresses as input would slow down the ML model prediction speed.
Instead, we extract the memory dependency by explicitly comparing the memory
access addresses of the current instruction with those of context instructions
and generate several {\em memory dependency flags} as input features.
For example, we compare their program counters (PCs) to identify if they fall
into the same instruction cache line.
This PC dependency flag presumably helps with fetch latency prediction as
instructions that share the same instruction cache line can be fetched
together.
Similarly, there are dependency flags to indicate if data accesses share the
same address, cache line, and page.

In addition, we introduce several features to capture the temporal
relationship between instructions.
Particularly, we include the number of cycles it has stayed in the processor
(i.e., residence latency), how long it takes to complete execution (i.e.,
execution latency), and the memory store latency (i.e., store latency) if
applicable.
The latter two are provided by the ML model output, which will be introduced
shortly.
The latency of context instructions is useful to predict the latency of the
current one.
For example, when an instruction follows a mispredicted branch, its fetch
latency is decided by the execution latency of the branch.


\subhead{Modeling History Context through Simplified Simulation.}
History context reflects the hardware states that depend on long-term
historical events, and it includes caches, TLBs, and branch predictors.
It is impractical to either capture the history context within the ML model or
directly have it as the input because it includes a huge amount of information.
Considering a 2MB cache with 64B cache lines as an example, we will need
$\sim$5B per cache line to store its address tag, etc.
Totally, a 2MB cache requires storing at least $2\text{MB} \div 64\text{B}
\times 5\text{B} = 160\text{KB}$ of information to simulate it accurately.
The total history-context-related information is much larger given all history
context.
It is prohibitively expensive and inefficient to let the ML model memorize such
large amounts of information.

Fortunately, the majority of history context impacts can be captured using a
small number of intermediate results.
For a memory access, the cache/TLB level in which it gets hit roughly
determines its latency.
Similarly, whether or not a branch target is predicted correctly determines the
impact of a branch prediction.

Therefore, we propose to simulate history context components explicitly to
obtain these intermediate results (i.e., {\em history context simulation}),
which are passed to the ML model as input features.
History context simulation greatly alleviates the burden on ML models.
As shown in the last row of Table \ref{tbl:features}, a branch misprediction
flag is obtained for a branch instruction.
An {\em access level} feature is used for each memory access to indicate which
level of the cache/TLB hierarchy satisfies the request.
All instructions require fetch access and fetch table walking levels, and
load/store instructions need data access and data table walking levels, e.g., a
load request that hits in the L2 cache has a level of 2.
The numbers of cache writebacks generated are also included in input features
to capture their impacts.

Note that obtaining these intermediate results mostly involves table lookups
(e.g., cache tag array; branch direction predictor).
Detailed structures, such as pipeline and miss status history register (MSHR),
are not needed in the history context simulation.
The impacts of these structures are captured by the ML model in \simnet{}.
Therefore, the history context simulation is lightweight and has negligible
impact on the overall performance.


\subhead{ML Model Input Summary.}
In total, each context instruction has 50 input features, and the
to-be-predicted instruction has 47 input features.
For alignment purposes, we pad the to-be-predicted instruction features with
three zeros, to have an equal number of 50 features.
Together, the ML model takes $50 \times (\text{\# context instructions} + 1)$
features as input.
While it is common to adopt one-hot encoding for individual input features, we
choose not to do so to favor smaller input size and faster prediction speed.

%

\subhead{ML Model Output.}
In Figure \ref{fig:mlsim}, the ML model is designed to predict three types of
latencies per instruction: fetch, execution, and store.
Fetch latency represents how long an instruction needs to wait to enter the
processor after the previous instruction is fetched.
It is affected by both its instruction fetch request and context instructions
(e.g., when it follows a mispredicted branch).
Execution latency represents the time interval from when an instruction is
fetched to when it finishes execution and is ready to retire from the reorder
buffer (ROB).
Note it is different from the ROB retire latency because the ROB retires
instructions in order.
For store instructions, they write memory after being retired from the ROB.
The store latency is used to represent the latency from when a store
instruction is fetched to when it completes memory write (i.e., when it is
ready to retire from the store queue (SQ)).
Section \ref{sect:sim:insttosim} will introduce how these latencies are used in
\simnet{} simulation.

\subsecspace
\subsection{Neural Network Architecture}
\label{sect:ml:arch}
\subsecspace
%
%

Given the input and output, we train various ML models to learn their
connections and capture the architectural impact.

\subhead{Sequence-Oriented Models.}
The ML model input includes a sequence of instructions (i.e., to-be-predicted
instruction and context instructions), similar to word sequences in the case of
natural language processing (NLP).
Therefore, a natural option is to apply models designed to process sequences,
such as recurrent neural networks \cite{rnn}, LSTM \cite{lstm}, and Transformer
\cite{transformer}, for instruction latency prediction.
Ithemal \cite{mendis2019ithemal} follows this strategy and adopts LSTM to
predict basic block latency.
The main drawback of these models is they are more computational intensive,
resulting in low simulation throughput.

\subhead{Deep Convolutional Neural Network (CNN) Models.}
Deep CNN models have shown great success in computer vision \cite{alexnet,
resnet, ICML19:EfficientNet}, where convolution kernels learn and recognize the
spatial relationship between pixels.
In our instruction latency prediction setting, convolution can help learn the
relationship among input instructions.
CNNs are less computational demanding than sequence-oriented models and fully
connected networks.
Another benefit of CNN is it eases the training of deeper networks because
significantly less parameters need to be learned.
As will be demonstrated in Section \ref{sect:ml:accuracy}, we choose CNNs for
\simnet{}'s instruction latency predictors due to their prediction accuracy and
computation overhead advantages.

\begin{figure}[t]
\centering
\includegraphics[width=0.45\linewidth]{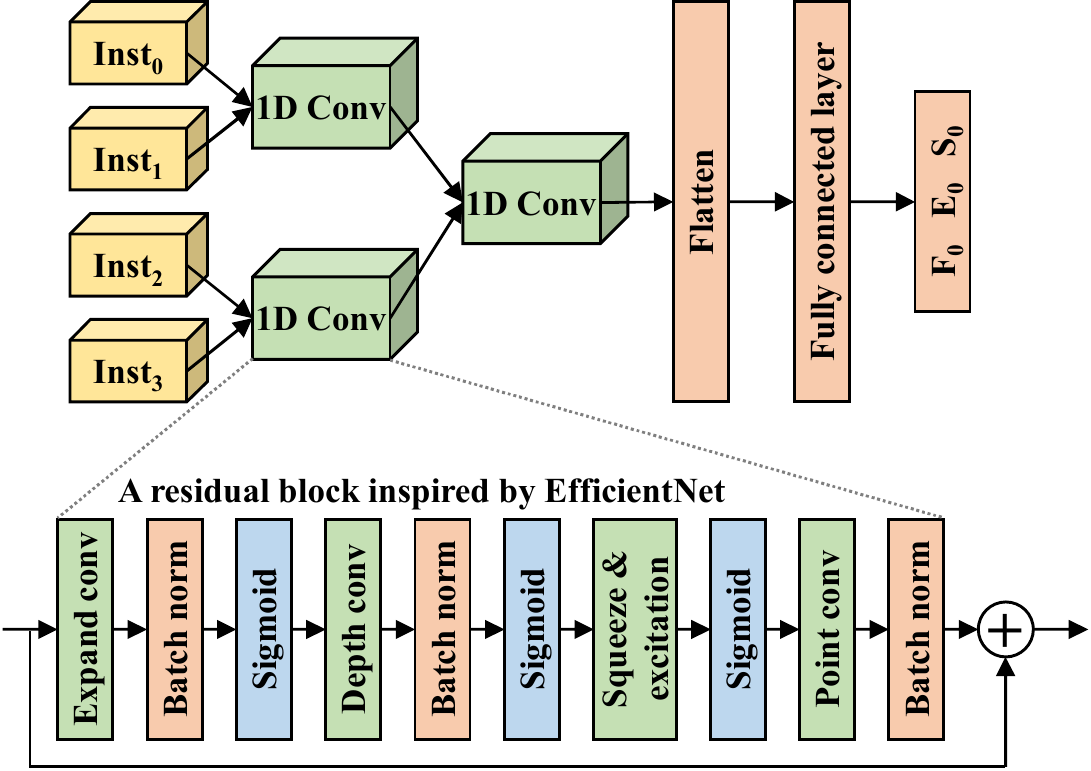}
\caption{Convolutional neural network architecture illustration.}
\label{fig:nn}
\end{figure}

Figure \ref{fig:nn} illustrates the proposed CNN architecture.
$\text{Inst}_0$ represents the instruction to be predicted, and the ML model
outputs $F_0$, $E_0$, and $S_0$, which are its predicted fetch, execution, and
store latencies, respectively.
Without loss of generality, Figure \ref{fig:nn} shows three context
instructions, $\text{Inst}_{1,2,3}$.

We organize input instructions in a one-dimensional (1D) array by their
execution order and have their features as channels, per CNN terminology.
As introduced in Section \ref{sect:mot:io}, every instruction includes 50
features.
Using computer vision as an analogy, instructions correspond to pixels, except
they are 1D instead of two-dimensional, and instruction features correspond to
pixel color channels.
This input organization facilitates convolutional operations to reason the
relationship between instructions.
Again, it is analogous to reasoning the shape composed by pixels in computer
vision.

We organize the convolutional layers in a hierarchical way, where the first
layer captures the relationship between temporally adjacent instructions, and
subsequent layers integrate the impact of further away instructions.
In Figure \ref{fig:nn}, the impact of $\text{Inst}_1$ to $\text{Inst}_0$ is
captured in the first layer, and the impact of $\text{Inst}_2$ and
$\text{Inst}_3$ is incorporated in the second layer.
This hierarchical design prioritizes the impact of temporally closer context
instructions while penalizing the influence of more distant instructions, and
real processors follow the same principle.
For instance, if a source register of $\text{Inst}_0$ is the destination
register of both $\text{Inst}_1$ and $\text{Inst}_3$, $\text{Inst}_0$ only
has to wait for $\text{Inst}_1$ where a true read after write dependency exists.

In our default design, each convolution layer includes a convolution operation
followed by an activation operation.
An alternative is to use a residual block as shown at the bottom of Figure
\ref{fig:nn}, which facilitates to increase the depth of CNNs \cite{resnet}.
In this work, we design a residual block architecture inspired by the
state-of-the-art image recognition model, EfficientNet
\cite{ICML19:EfficientNet, CVPR19:MnasNet}.

The output of the last convolutional layer is flattened then used as the input
of two fully connected layers.
At the end, the model outputs the predicted fetch, execution, and store
latencies of $\text{Inst}_0$.
We adopt the commonly used rectified linear unit (ReLU) as the activation
function of both the convolutional and fully connected layers.

Empirically, we find the following CNN design principles work well for
instruction latency prediction.
First, the inputs of different convolutional operations have no overlap in
contrast to computer vision CNNs.
For example, we do not convolve $\text{Inst}_1$ and $\text{Inst}_2$ in Figure
\ref{fig:nn}.
In this way, the impact of a context instruction is integrated only once.
Second, a convolution kernel size of 2 is always used to account for only two
adjacent inputs, which reduces the complexity.
Combined with the first principle, it means all convolutional layers have the
uniform kernel and stride size, 2.
Our experiments demonstrate that these principles work well across different
architecture configurations.
While an extensive neural architecture search \cite{elsken2019neural} could
potentially find better architectures, it also means significant searching
overhead and we leave it for future work.

%
%

\subhead{From Output to Latency.}
There are two ways to convert the ML model output to the
latency prediction results.
In a {\em regression} model, the model output is directly used as the predicted latency.
One inherited issue for the regression latency prediction model is its
inability to distinguish between small latency differences.
The impact may be minor when a latency of 1000 cycles is predicted to be 1001
cycles, but the error could be significant for small latencies (e.g., 0 cycle
predicted to be 1).
Because the fetch latency is 0 or 1 cycle in most cases, this drawback is
particularly critical for its prediction.

A {\em classification} model could help to better distinguish between close latency
values, where every latency value corresponds to a class, and the ML model
predicts which class has the largest probability.
However, because the latency could be up to several thousands of cycles, a pure
classification scheme will significantly increase the output size and, thus, the
computational overhead.
Another problem of a pure classification scheme is it is difficult to train
such a model because large latency samples appear less frequently in the
training set.

As it is quite expensive to have a class for each possible latency value, we
propose a {\em hybrid} scheme which uses classification for latency that
appears frequently and regression for others.
Naturally, small latencies appear more frequently.
Taking the fetch latency prediction as an example, we classify them into 10
classes in the hybrid scheme.
Cycles 0 to 8 have dedicated classes ($c_0, ...,c_8$), while another
class is used to represent cycles that are larger than 8 ($c_{>8}$).
The proposed model outputs the probability of each class.
It also outputs a direct prediction result $l$ as in the regression model.
On a prediction, we first check which class has the largest probability.
If it is one among $c_0, ...,c_8$, the corresponding latency is predicted.
Otherwise, $l$ is used as the predicted latency.
Similar procedures are used to predict execution and store latencies.


\subsecspace
\subsection{Dataset and Training}
\label{sect:ml:data}
\subsecspace

\subhead{Data Acquisition.}
Due to their data-driven nature, acquiring a sufficient training dataset is
necessary for the success of ML-based approaches.
Fortunately, it is convenient to engage existing simulator infrastructures to
acquire a dataset for standard {\em supervised} training.

We modify gem5 \cite{gem5} to dump instruction execution traces, which then are
used to generate ML training/validation/testing datasets.
In the modified gem5, each instruction is assigned with three timestamps to
record its respective fetch, execution, and store latencies.
While the fetch latency stamp is updated in the instruction fetch unit, the
execution and store latency stamps are updated in the ROB and SQ, respectively.
After all latencies of an instruction are recorded, gem5 dumps it to a trace
file.

The instruction traces output by gem5 require several steps of processing
before they can be used for ML.
First, for each instruction, we find and associate its context instructions
based on the timestamps to form a \emph{sample}.
Second, many samples may be alike because the same scenarios can appear
repeatedly during the execution of benchmarks.
We eliminate such duplication to reduce the dataset.
Finally, we convert the dataset to the format used by the ML framework.

\begin{table}[t]
\scriptsize
\centering
\begin{tabularx}{\linewidth}{|l|X|X|}
\hline
{\bf Parameter} & {\bf Default O3CPU} & {\bf A64FX} \\
\hline
\hline
Core & 3-wide fetch, 8-wide out-of-order issue/commit, bi-mode branch predictor, 32-entry IQ, 40-entry ROB, 16-entry LQ, 16-entry SQ & 8-wide fetch, 4-wide out-of-order issue/commit, bi-mode branch predictor, 48-entry IQ, 128-entry ROB, 40-entry LQ, 24-entry SQ \\
\hline
L1 ICache & 48KB, 3-way, LRU, 4 MSHRs & 64KB, 4-way, LRU, 8 MSHRs \\
\hline
L1 DCache & 32KB, 2-way, LRU, 16 MSHRs, 5-cycle latency & 64KB, 4-way, LRU, 21 MSHRs, 8-cycle latency, 8-degree stride prefetcher \\
\hline
I/DMMU & 2-stage TLBs, 1KB 8-way TLB caches with 6 MSHRs & 2-stage TLBs, 1KB 4-way TLB caches with 6 MSHRs \\
\hline
L2 Cache & 1MB, 16-way, LRU, 32 MSHRs, 29-cycle latency & 8MB 16-way, LRU, 64 MSHRs, 111-cycle latency \\
\hline
\end{tabularx}
\caption{Simulated processor configurations.}
\label{tbl:arch}
\end{table}

Table \ref{tbl:arch} shows the processor configurations that ML models learn
from.
The default O3CPU resembles a classic superscalar CPU.
We also train models to learn the Fujitsu A64FX CPU deployed in the current
top-ranked supercomputer, Fugaku \cite{HC18:A64FX, SC20:A64FX}, which
represents a state-of-the-art CPU.
We obtain the official gem5 configurations of A64FX at
\cite{kodama2019evaluation}, which is verified to have an average simulation
error of 6.6\% against the real processor.
Both simulated processors support the ARMv8 instruction set architecture (ISA),
and benchmarks are compiled using gcc 8.2.0 under the O3 optimization level.
The full system simulation mode of gem5 is employed with Linux kernel 4.15.
We use the default O3CPU configuration for most of our experiments, while
Section \ref{sect:eval:accuracy} will present the results for the A64FX
configuration.
Under the default O3CPU configuration, there are, at most, 110 context
instructions.
Therefore, the ML model input has $50 \times (1+110) = 5550$ features.

ARMv8 is a representative 64-bit reduced instruction set computer (RISC).
It includes various integer, floating-point, branch/jump, load/store,
vectorized floating-point/integer, Boolean logic instructions, etc.
A trained \simnet{} model can predict the latency of all these instructions.
We expect \simnet{} will be able to support future ISA extensions.

\simnet{} can also support other ISAs including complex instruction set
computers (CISCs) such as x86.
To directly predict the performance of CISC instructions, more input features
(e.g., multiple data access levels) are required because they show more complex
behaviors such as multiple memory accesses.
Another possible approach to support CISCs is to decompose CISC instructions
into RISC like macro instructions, similar to what contemporary CISC CPUs do.
In this way, \simnet{} can be used to predict the latency of simpler RISC
instructions, similar to ARM ones.

\subhead{Benchmark.}
Theoretically, any program can be run on the modified gem5 to collect the ML
dataset, and we can acquire an unlimited amount of data.
We choose to use the SPEC CPU 2017 \cite{bucek2018spec} benchmark suite in this
paper because it is widely used in computer architecture simulation and
includes a wide range of applications, which should lead to a sufficient
coverage of instruction execution scenarios.
We select the first four SPEC CPU 2017 benchmarks to generate the ML
training/validation/testing dataset, which are shown in Table
\ref{tbl:benchmarks}.
The default test workloads are used for these four benchmarks, and one billion
instructions are simulated from the beginning for each benchmark to collect the
ML dataset.
Totally, we obtain a dataset with 71 million samples, among which roughly 90\%
of them are dedicated for training, 5\% for validation, and 5\% for testing.


As will be introduced in Section \ref{sect:eval}, we use the reference
workloads to verify the simulation accuracy of all 25 SPEC CPU 2017 benchmarks.
The facts that 21 benchmarks of them do not appear in the ML dataset and the
simulation accuracy is evaluated on different input workloads, allow us to
evaluate the generalizability of \simnet{}.


\begin{table}[t]
\scriptsize
\centering
\begin{tabular}{|c|c|c|}
\hline
{\bf Type} & {\bf ML} & {\bf Simulation} \\
\hline
\hline
INT & \minitab[c]{perlbench,\\ gcc} & \minitab[c]{mcf, omnetpp, xalancbmk, x264, deepsjeng,\\ leela, exchange2, xz, specrand\_i} \\
\hline
FP & \minitab[c]{bwaves,\\ namd} & \minitab[c]{cactuBSSN, parest, povray, lbm, wrf, blender,\\ cam4, imagick, nab, fotonik3d, roms, specrand\_f} \\
\hline
\end{tabular}
\caption{Benchmarks for ML and simulation.}
\label{tbl:benchmarks}
\end{table}


\subhead{Training.}
We use the standard gradient-based optimization to train various models.
Let $\{ (x_i,y_i)\}_{i=1}^{n}$ represent the set of input and output pairs in a
training set of $n$ samples.
Let $f_{\theta}$ represent a to-be-trained model with parameters $\theta$, and
our goal is to find a particular $\theta$ that minimizes the training loss
$J(\theta) = \frac{1}{n}\sum\limits_{i=1}^{n}L( f_{\theta}(x_i),y_i).$
When training the regression output, $L$ is the squared-error loss function.
When training the classification output, $L$ is the cross-entropy loss
function.

Our training code is built upon PyTorch 1.7.0 \cite{pytorch}.
The objective function $J$ is minimized using the Adam optimizer
\cite{kingma2015adam}.
We use a learning rate of 0.001 and no weight decay or momentum.
Every model is trained for 200 epochs, and the validation set is used to select
the model with the lowest loss.
No hyperparameter tuning is performed when training for different architecture
configurations to avoid extensive hyperparameter search overhead.
Our ML training hardware platform is an NVIDIA DGX A100 system \cite{dgxa100}.
It includes eight NVIDIA A100 GPUs connected through NVLink 3.0 and NVSwitch,
and each is equipped with 40GB HBM that supports 1.5 TB/sec peak bandwidth.
Tensor cores in an A100 GPU enable a peak performance of 156 TFlops for Tensor
Float 32 operations.
The DGX A100 system's high computing and memory throughputs make it ideal for
ML training and inference.
Depending on its complexity, training a model takes $18 \sim 75$ hours on this
machine.
Section \ref{sect:eval:overhead} will discuss the training overhead.

\begin{table*}[t]
\scriptsize
\centering
\begin{tabular}{|c|c|c|c||c|c|c||c|c|c|}
    \hline
    \multirow{2}{*}{Approach} & \multirow{2}{*}{ML model} & \multirow{2}{*}{Output} & \multirow{2}{*}{\minitab[c]{Computation\\intensity (MFlops)}} & \multicolumn{3}{c||}{Instruction prediction error} & \multicolumn{3}{c|}{Benchmark simulation error} \\
    \cline{5-10}
    & & & & Fetch & Execution & Store & train avg. & sim. avg. & all avg. \\
    \hline
    \multirow{8}{*}{\simnet{}} & FC2    & reg & 5.7  & 89\% & 11\% & 71\% & 82\% & 57\% & 61\% \\
    \cline{2-10}
    & FC3 & reg & 6.7  & 32\% & 6.1\% & 4.1\% & 14\% & 20\% & 19\% \\
    \cline{2-10}
    & C1 & reg & 4.0  & 55\% & 9.3\% & 8.5\% & 19\% & 31\% & 29\% \\
    \cline{2-10}
    & C3 & reg & 8.1  & 35\% & 6.4\% & 17\% & 4.6\% & 9.0\% & 8.3\% \\
    \cline{2-10}
    & C3 & hyb & 8.1  & 2.7\% & 3.4\% & 1.0\% & 2.7\% & 12\% & 10\% \\
    \cline{2-10}
    & RB7 & hyb & 93 & 1.7\% & 1.8\% & 0.6\% & 5.7\% & 5.5\% & 5.6\% \\
    \cline{2-10}
    & LSTM2 & hyb & 119 & 6.5\% & 6.0\% & 1.6\% & 7.3\% & 7.9\% & 7.8\% \\
    \cline{2-10}
    & TX6 & hyb & 1185 & 6.4\% & 4.2\% & 1.1\% & 7.9\% & 9.6\% & 9.3\% \\
    \hline
    \multirow{2}{*}{Ithemal} & LSTM2 & \multirow{2}{*}{N/A} & 216 & 64\% & 17\% & 37\% & 20\% & 27\% & 26\% \\
    \cline{2-2}\cline{4-10}
    & LSTM4 & & 487 & 68\% & 14\% & 74\% & 15\% & 16\% & 16\% \\
    \hline
\end{tabular}
\caption{Instruction latency prediction and program simulation accuracy of
various ML models. Output indicates if it is a regression model (reg) or hybrid
model with classification (hyb). Computation intensity measures the number of
million floating point multiplications (MFlops) required for one inference.}
\label{tbl:modelaccuracy}
\end{table*}

\subsecspace
\subsection{Model Evaluation}
\label{sect:ml:accuracy}
\subsecspace

We evaluate an array of ML models for instruction latency prediction, and
the middle part of Table \ref{tbl:modelaccuracy} compares their prediction
accuracy.
We represent an ML model using a combination of letters and numbers, where the
prefix denotes the basic building block type and the suffix denotes the number
of layers.
Particularly, FC, C, RB, LSTM, and TX represent the fully connected
layer, the conventional convolutional layer, the residual block depicted at the
bottom of Figure \ref{fig:nn}, the standard LSTM block, and the Transformer
encoder layer \cite{transformer}, respectively.
For example, C3 is composed of three conventional convolutional layers.

The prediction error of each latency type is defined as follows for the $i$th
entry of testing dataset: $E = \frac{|f_\theta(x_i) - y_i|}{y_i + 1}$, where
$x_i$ is the input and $y_i$ is the expected output.
Note that we use $y_i + 1$ as the denominator instead of $y_i$ because the
$y_i$ of fetch and store latencies (e.g., non stores) is often 0.

Table \ref{tbl:modelaccuracy} affords several observations.
First, we note that the prediction error of CNNs improves with the number of
layers, which demonstrates the necessity of a {\bf deep} neural network.
Particularly, RB7 with residual blocks achieves the best accuracy, while the
simplest FC2 model's prediction error is an order of magnitude larger.

Second, the {\bf hybrid} scheme helps reduce prediction errors, from 35\% to
2.7\% for fetch latency's error under C3, while barely increasing the
computation complexity.
We notice that the hybrid C3 model makes correct fetch latency predictions in
95\% of cases.
In comparison, the regression C3 model predicts 65\% of fetch latency
correctly, which demonstrates that classification is helpful to predict
latencies with small values.

Third, Table \ref{tbl:modelaccuracy} also compares the computational overhead of
various models.
CNN models require $4 \sim 93$ millions of multiplications per
inference/prediction.
Different models represent different trade-off points between accuracy and
computation overhead.
Although they seem to be higher than that of traditional simulators, these
computations are performed very efficiently on modern accelerators, such as GPUs
and TPUs.
As a result, \simnet{} achieves significantly higher simulation throughputs as
well as better power efficiency, as will be shown in Section
\ref{sect:eval:par}.

Compared with CNN models, both LSTM and Transformer models show lower
prediction accuracy while incurring much larger computational overhead.
Transformer models are especially expensive due to the attention computation
\cite{transformer}.
Although there are spaces to improve their accuracy through the adoption of
deeper and wider networks,  doing so requires larger computation overhead.
This demonstrates that CNNs are more efficient at instruction latency
prediction under a constrained computation budget.

\subhead{Comparison with Ithemal.}
We also compare \simnet{} with Ithemal \cite{mendis2019ithemal}, a
state-of-the-art ML-based latency prediction approach.
It is designed to predict the latency of a basic block for processors with
ideal caches and branch predictors, i.e., all memory accesses hit in L1 caches,
and branch latency is not considered.
To make a meanful comparison, we enhance it with \simnet{} input features so
that it considers the impact of realistic caches and branch predictors and can
predict the latencies of instruction sequences that are much longer than basic
blocks.
The key difference between Ithemal and \simnet{} is that the former uses a fix
number of previous instructions as the input, while the latter explicitly
selects instructions that are active in the processor (i.e., context
instructions) as its input and excludes those that have retired.

We train two LSTM models using the Ithemal approach, and the last two rows of
Table \ref{tbl:modelaccuracy} show their results.
The LSTM4 model is similar to what Ithemal originally uses, and we also include
a 2-layer LSTM.
The same training dataset and process as \simnet{} is adopted to ensure
fairness.
We find LSTM4 does not always perform better than LSTM2 due to the fading
gradient problem that is common for deep LSTM training.

We observe that \simnet{}'s prediction errors are one order of magnitude lower
than those of Ithemal.
\simnet{} models also incur lower computation overhead.
These results demonstrate that explicitly constructing context instructions
significantly improves instruction latency prediction accuracy, which is a key
contribution of \simnet{}.
This conclusion is intuitive because the input can better reflect the processor
status when excluding retired instructions, which simplifies the job of ML
models and results in higher accuracy.
Note that the LSTM2 models of \simnet{} and Ithemal have the same architecture.
\simnet{}'s LSTM2 has a lower computation intensity because \simnet{} has less
instructions as input.
The fact that \simnet{}'s LSTM2 has significantly better prediction accuracy
than Ithemal's LSTM2 further demonstrates the effectiveness of \simnet{}.

\secspace
\section{ML-based Simulation}
\label{sect:sim}
\secspace

\begin{figure}[t]
\centering
\includegraphics[width=0.5\linewidth]{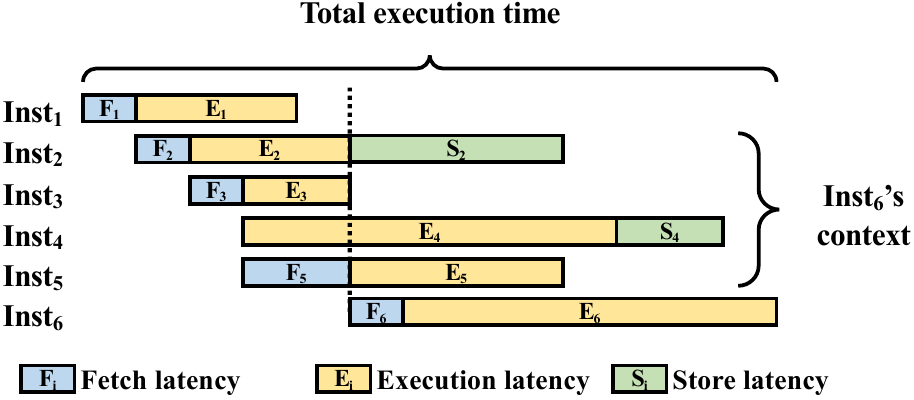}
\caption{From instruction latency to program execution time.}
\label{fig:lat-to-time}
\end{figure}

\subsecspace
\subsection{From Instruction Latency to Program Performance}
\label{sect:sim:insttosim}
\subsecspace

\simnet{} simulates the program performance using the ML-based instruction
latency predictor introduced in Section \ref{sect:ml}.
Figure \ref{fig:lat-to-time} illustrates how to calculate program execution
time using instruction latencies, leveraging the fact that instruction fetch
and instruction retire from ROB and SQ happen in order.
Note that the fetch latency could be 0 in cases when multiple instructions are
fetched together (e.g., $\text{Inst}_4$).
We observe the execution time $\mathcal{E}$ of a program can be computed as
\begin{equation}
\label{equ:time}
\mathcal{E} = (\sum_{i=1}^n F_i)+\Delta,
\end{equation}
where $n$ is the total number of simulated instructions, $F_i$ represents the
fetch latency of the $i$th instruction, and $\Delta$ is the amount of time from
when the last instruction is fetched to when all instructions exit the
processor.
When $n$ is large enough, the total execution time is dominated by the
accumulated fetch latencies, and $\Delta$ is negligible.
This equation lays down the foundation for the proposed instruction-centric
simulator.

\subsecspace
\subsection{Simulator Implementation}
\label{sect:sim:cxt}
\subsecspace

Based on Equation \ref{equ:time}, we develop a trace-driven simulator.
The simulator goes through every executed instruction instance to predict its
latency and outputs the program performance upon completion.
The modified gem5 is used to generate input traces,
which include instruction properties extracted by functional simulation, and
history context simulation results.

\subhead{Context Management.}
As introduced in Section \ref{sect:mot:io}, the ML predictor requires the features of context instructions as part of its
input.
Therefore, \simnet{} needs to keep track of context instructions.
For example, in Figure \ref{fig:lat-to-time}, when $\text{Inst}_6$ is about to
be fetched (vertical black-dotted line), $\text{Inst}_1$ has retired, and
$\text{Inst}_{2\sim5}$ are still in the processor based on their execution and
store latencies.
Therefore, $\text{Inst}_{2\sim5}$ are the context instructions of
$\text{Inst}_6$.
To this end, we employ two first-in-first-out (FIFO) queues, {\em
processor queue} and {\em memory write queue}, to keep track of context
instructions that stay in the processor and their features.
They roughly correspond to the ROB and SQ in an out-of-order processor but are
not exactly the same.
The two major differences are that the processor queue includes instructions in
the frontend, while ROB does not, and a store instruction enters the memory
write queue after it retires from the processor queue.

After the simulator reads one instruction from the input trace, the ML
predictor is invoked to predict its latency.
Then, it enters the processor queue with the residence latency initialized to 0.
When it retires from the processor queue is determined based on its predicted
execution latency and other simulation constraints (e.g., it must obey the
in-order retirement and retire bandwidth).
A non-store instruction exits the simulator when it retires from the processor
queue.
For a store instruction, it will enter the memory write queue.
Similarly, when an instruction retires from the memory write queue is decided
based on its predicted store latency, and it will exit the processor at that
time.
Much like real processors, the retire bandwidth of a processor queue is set
according to that of the ROB, and the memory write queue can retire any number of
instructions from its tail.


\subhead{Clock Management.}
The simulator employs {\tt curTick} to record the total number of simulation
cycles, which is updated whenever a prediction completes.
When the predicted fetch latency is larger than 0, it is added to {\tt curTick}
so that the counter always points to the time when the current instruction
enters the processor.
In this case, we also increase the residence latency of all context
instructions by the predicted fetch latency to update the time that they have
remained in the processor.
When the residence latency of an instruction is larger than its execution
latency, it is ready to retire from the processor queue.
Similarly, an instruction is ready to retire from the memory write queue when
its residence latency exceeds its store latency.

After the last instruction in the input trace is predicted, we continue
advancing {\tt curTick} until all instructions retire from the simulator.
The final value of {\tt curTick} represents the total execution time of the
program, which is exactly the same as Equation \ref{equ:time}.

\subsecspace
\subsection{GPU-accelerated Parallel Simulation}
\label{sect:sim:accel}
\subsecspace

In our ML-based instruction latency predictor, the latency of an instruction
depends on the predicted latencies of previous instructions, i.e., the latency
prediction of adjacent instructions is inherently sequential.
This restriction limits the sequential simulation speed and computational
resource utilization.
As a result, a sequential implementation of \simnet{} runs at a throughput of
$\sim$ 1k instructions per second, and it can only leverage a very small
fraction of modern GPU's computing power.
To improve the simulation throughput and resource utilization, we seek to
extract parallelism.

\begin{figure}[t]
    \centering
    \includegraphics[width=0.5\linewidth]{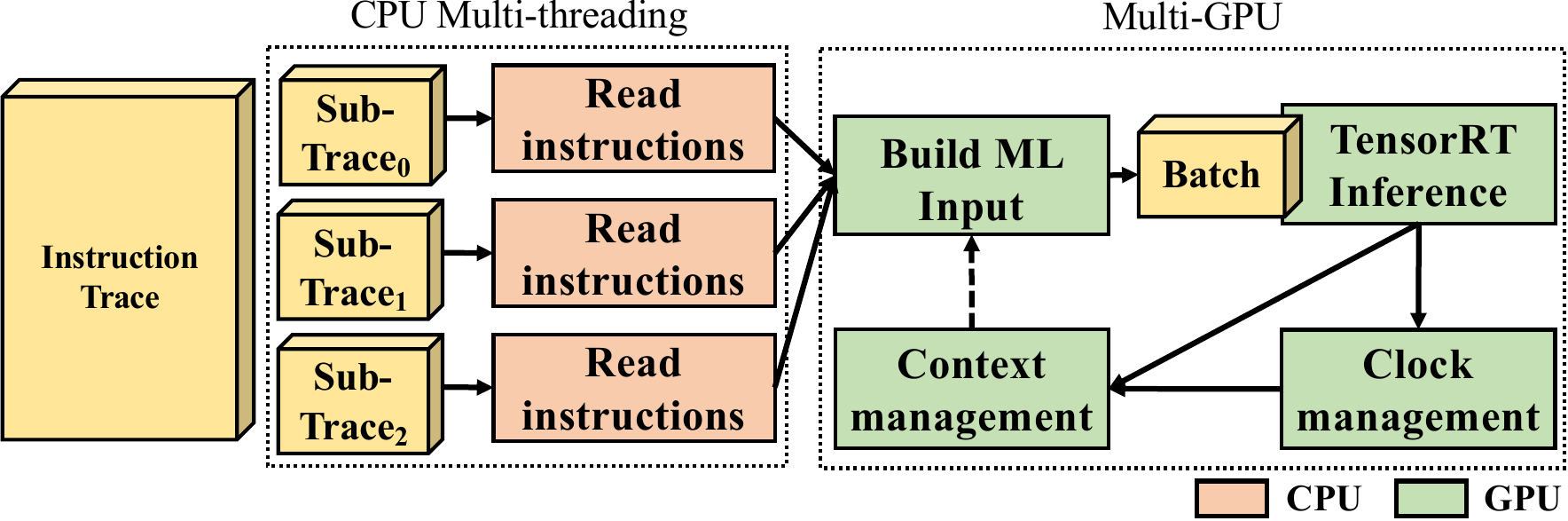}
    \caption{Parallel simulation framework.}
    \label{fig:batch}
\end{figure}

\subhead{Parallel Simulation of Sub-traces.}
The primary idea is to break down the input instruction trace into multiple,
equally sized continuous sub-traces and simulate sub-traces independently in
parallel.
The instructions within each sub-trace are simulated sequentially to preserve
the instruction dependency within the specific sub-trace.
The drawback of this approach is that extra simulation errors are introduced
when simulating earlier instructions of a sub-trace due to inaccurate or
missing contexts.
Section \ref{sect:eval:par} will show that such accuracy loss is
negligible when each sub-trace is large enough.
Figure~\ref{fig:batch} shows the overview of parallel simulation.
The design leverages CPU multi-threading to partition the input trace and
transfer sub-trace instructions to the GPU memory.
The remaining work is done by GPUs to capitalize on their high computational
capacity and reduce communications between CPUs and GPUs.


\subhead{GPU Acceleration.}
Both context and clock management are implemented on GPUs, and each sub-trace
has separate copies of them.
Particularly, each sub-trace has its own processor queue and memory write
queue, as well as a {\tt curTick} counter to record its number of simulated
cycles.
After the ML model input is built independently for each sub-trace, we combine
them into a single input to allow GPU-batched inferences.
This process repeats until all instructions in a sub-trace are simulated.
After all sub-traces complete their simulation, we sum up their {\tt curTick}s to
get the total execution time.
For ML model inferences, we use TensorRT \cite{tensorrt}, developed by
NVIDIA for high-performance GPU deep learning inferences.
It optimizes GPU memory allocations and supports reduced precision inferences.
We adopt the TF32 and FP16 formats for ML inferences in this paper, and expect
\simnet{} can benefit from the use of lower precisions when their supports
become more mature in TensorRT.
Compared with PyTorch, TensorRT provides roughly $3\times$ speedup.
In addition, this design can be scaled to multiple GPUs, where each GPU is responsible for a
fraction of sub-traces. No inter-GPU communication is required during the simulation process. 
Section~\ref{sect:eval:par} will offer a detailed evaluation of simulation
throughput.




\secspace
\section{Evaluation}
\label{sect:eval}
\secspace


\subsecspace
\subsection{Simulation Accuracy Validation}
\label{sect:eval:accuracy}
\subsecspace


\begin{figure*}
  \centering
  \includegraphics[width=\linewidth]{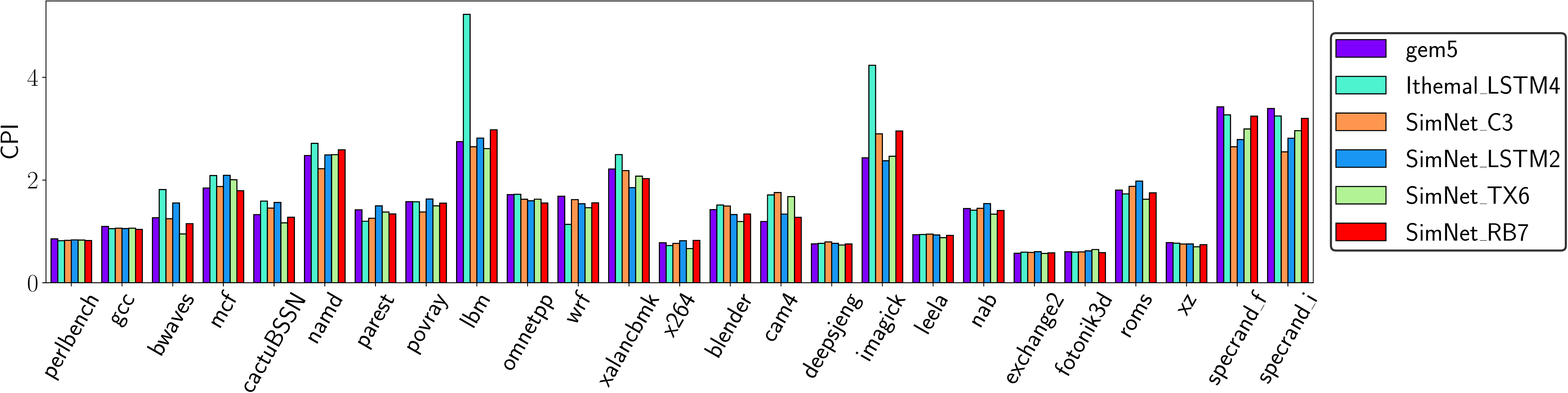}
  \caption{Simulated benchmark CPIs for various approaches.}
  \label{fig:accuracy}
\end{figure*}

\subhead{Benchmark Simulation Accuracy.}
We conduct simulation experiments on our training platform: the NVIDIA DGX
A100 system equipped with eight A100 GPUs and an AMD EPYC 7742 64-core CPU.
We simulate all 25 SPEC CPU 2017 SPECrate benchmarks using the reference
workload.
For each benchmark, SimPoint \cite{ASPLOS02:SimPoint} is used to select a
representative sample of 100 million instructions.

The right side of Table \ref{tbl:modelaccuracy} illustrates the simulation
errors of various models compared with gem5.
We use the absolute value of normalized cycle per instruction (CPI) difference
to measure the simulation error for each benchmark: $|CPI_\text{\simnet} /
CPI_\text{gem5} - 1| \times 100\%$.
Although models with lower instruction prediction errors have lower simulation
errors in most cases, it is not always true.
The reason is because previous prediction results are used to construct the
input of latter predictions through the instruction context, which leads to a
more complicated relationship between the predictor's and simulator's accuracy
as will be discussed later.

Table \ref{tbl:modelaccuracy} shows the average simulation errors across three
benchmark sets: benchmarks used in ML training (i.e., 4 ML benchmarks in Table
\ref{tbl:benchmarks}), benchmarks not used in ML training (i.e., 21 simulation
benchmarks in Table \ref{tbl:benchmarks}), and all of them.
Note that for benchmarks used in training, different input workloads (test vs.
reference) and simulation segments (beginning vs. SimPoint selected) are used
in their simulation.
We observe that the average errors of simulation workloads are not necessarily
larger than those of training benchmarks, and the formers are smaller for
several models.
It demonstrates \simnet{}'s ability to simulate unseen benchmarks.
\simnet{}'s generalizability roots in the fact that its predictor is trained at
the instruction level.

Among \simnet{} models, the deepest CNN model RB7 achieves the lowest average
simulation error of 5.6\%.
The shallower CNN model C3 also achieves good accuracy with a significantly low
computation cost.
Therefore, we focus on C3 and RB7 in the following experiments.
On the other hand, LSTM and Transformer models achieve comparable simulation
accuracy at a cost of one order of magnitude more computation overhead.
Compared with Ithemal models, \simnet{} ones have significantly lower errors,
which again demonstrates \simnet{}'s effectiveness by constructing context
explicitly.

Figure \ref{fig:accuracy} further compares the simulated CPIs of gem5, the most
accurate Ithemal model LSTM4, and representative \simnet{} models per
benchmark.
While Ithemal incurs significant errors for several benchmarks, \simnet{}
models accurately simulates most benchmarks whose CPIs spread across a wide
spectrum.
Among them, RB7 achieves the best simulation accuracy where only 1 out of 25
benchmarks has an absolute error $> 10\%$ (22\% for {\tt imagick}).

\begin{figure*}[t]
\centering
%
%
\begin{minipage}[t]{\textwidth}
\centerline{
  \includegraphics[width=0.7\textwidth]{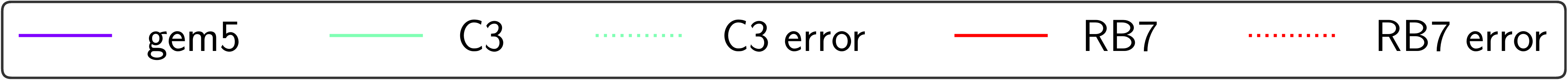}
}
\end{minipage}%

\begin{minipage}[t]{\textwidth}
\centerline{
  \includegraphics[width=0.2\textwidth]{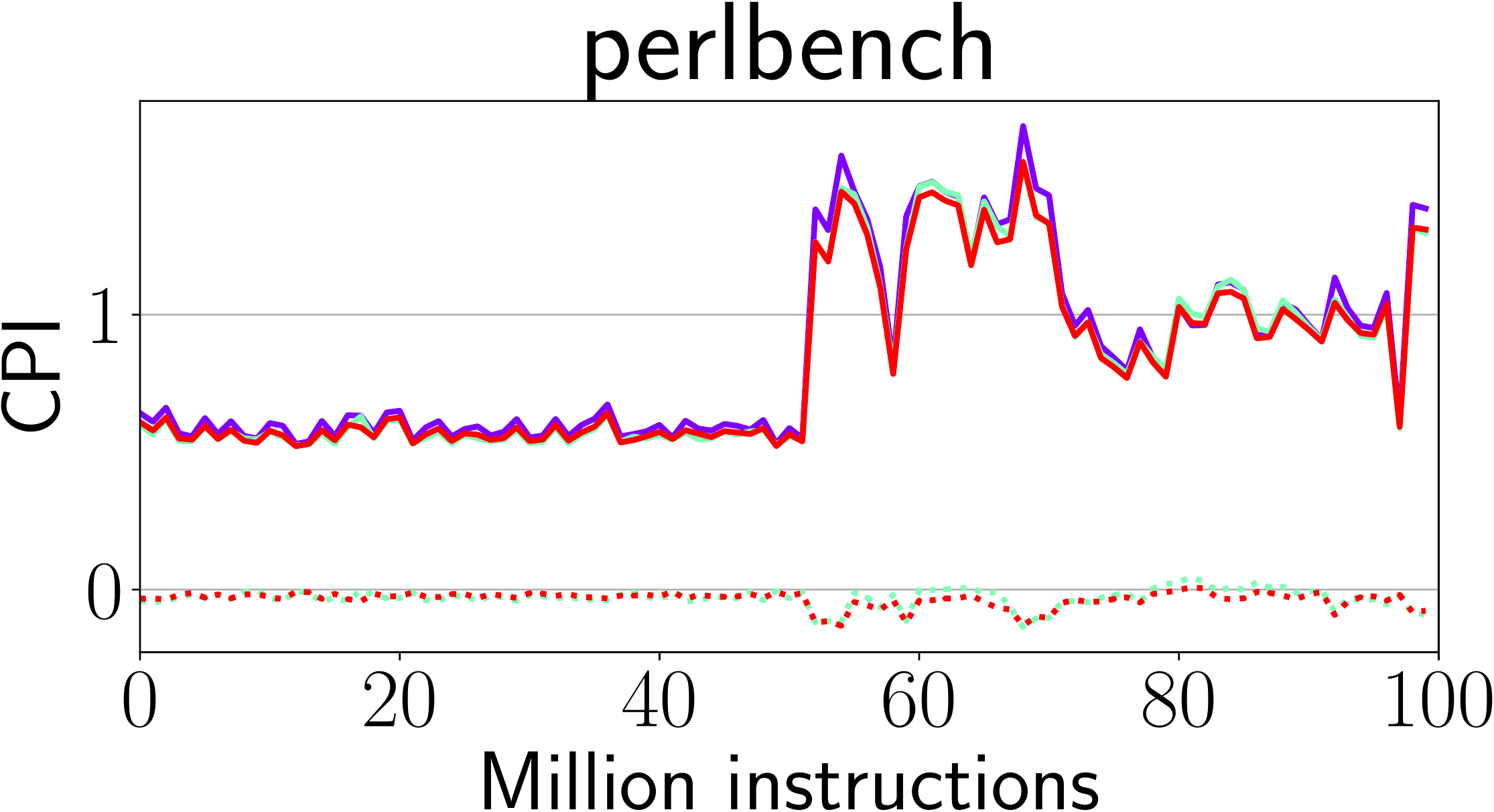}
  \hfil
  \includegraphics[width=0.2\textwidth]{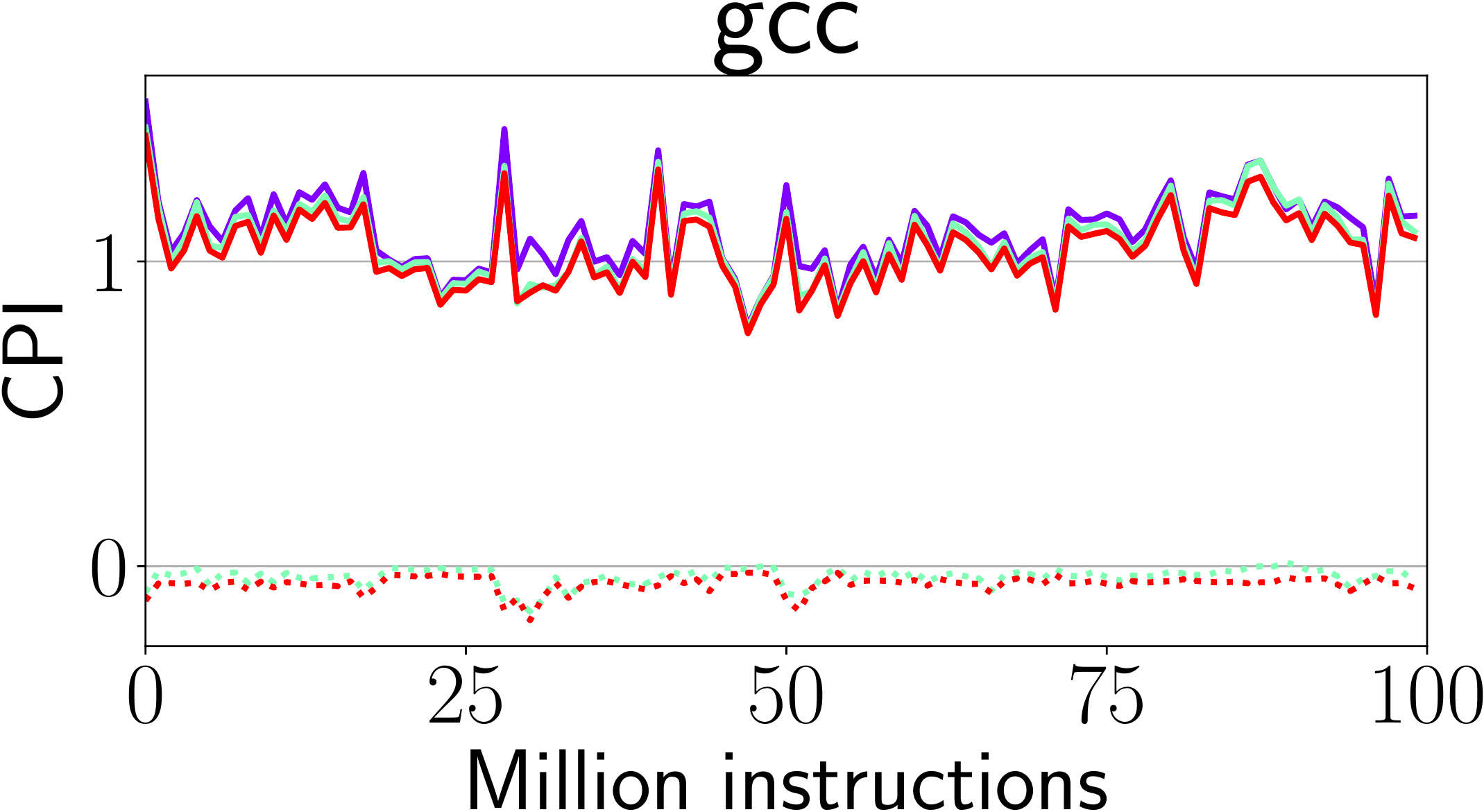}
  \hfil
  \includegraphics[width=0.2\textwidth]{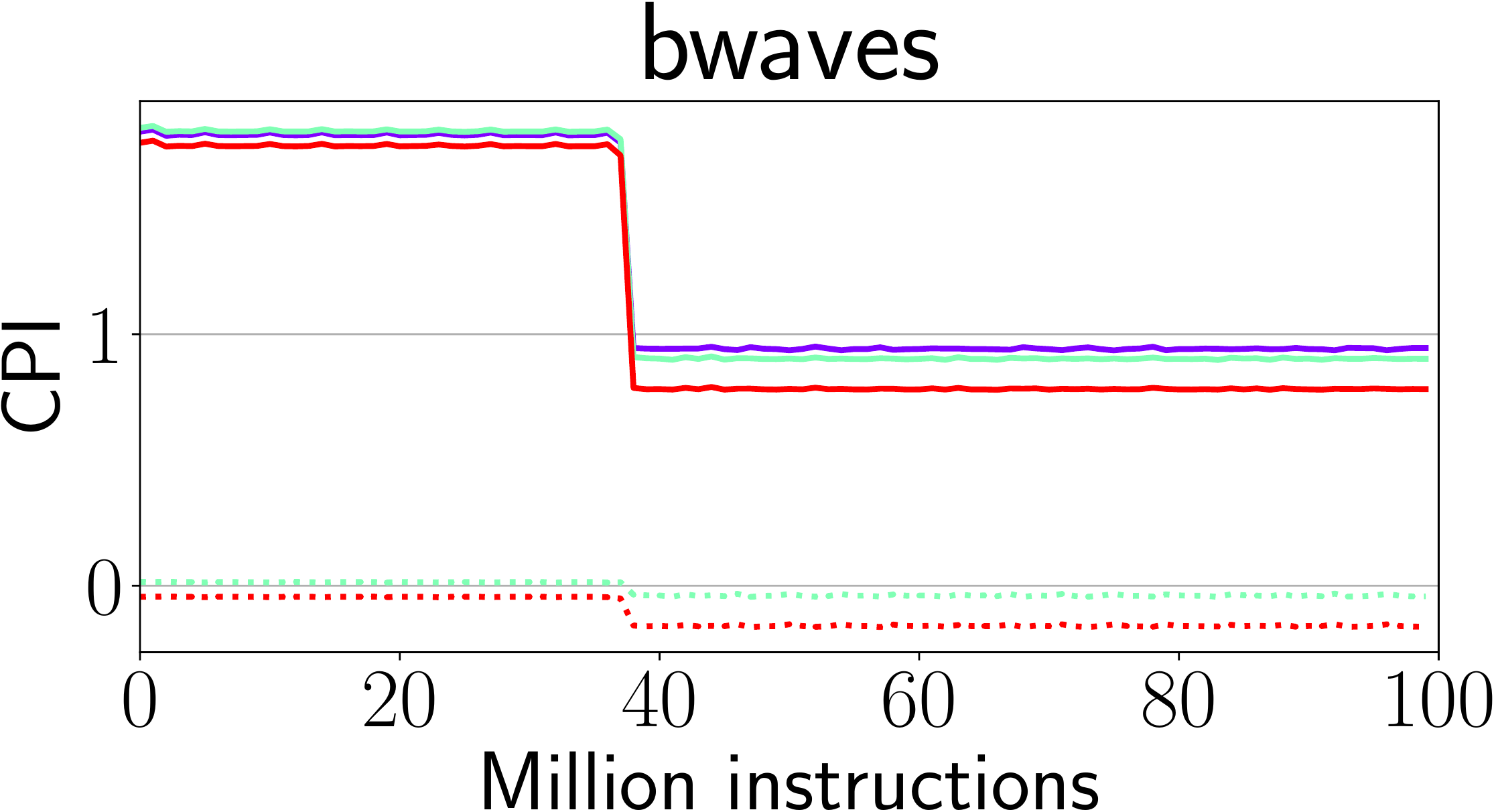}
  \hfil
  \includegraphics[width=0.2\textwidth]{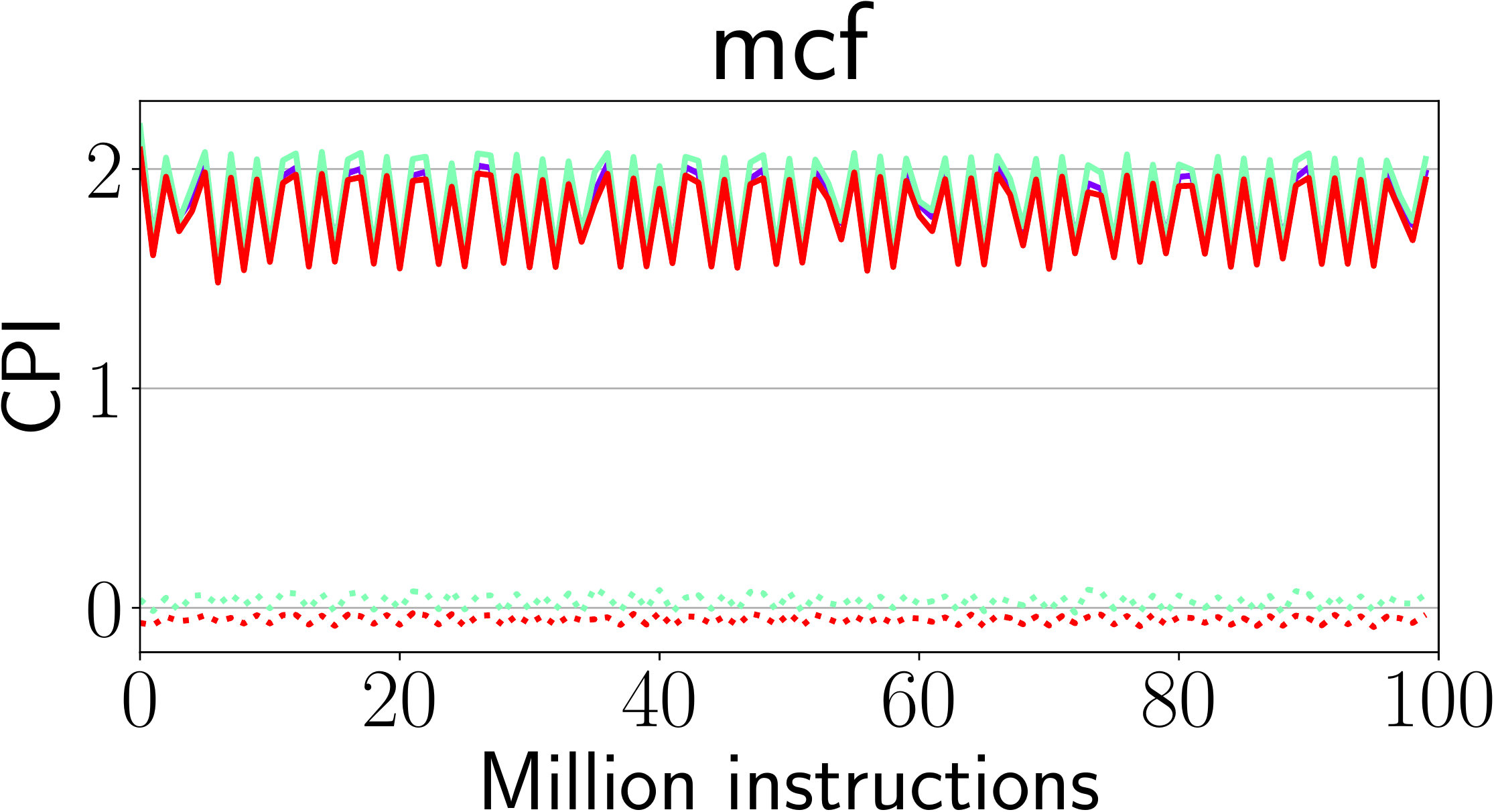}
  \hfil
  \includegraphics[width=0.2\textwidth]{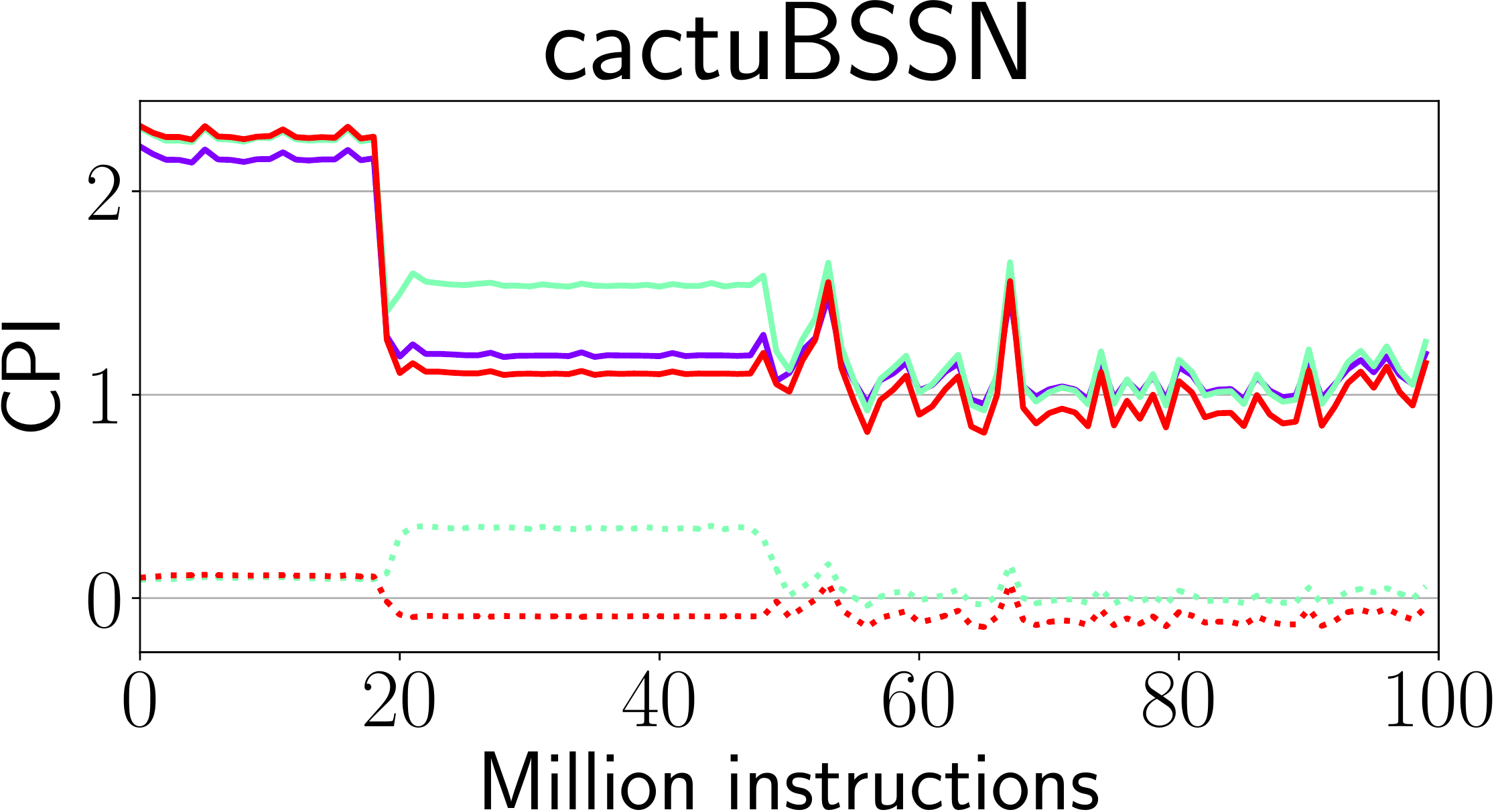}
}
\end{minipage}%

\begin{minipage}[t]{\textwidth}
\centerline{
  \includegraphics[width=0.2\textwidth]{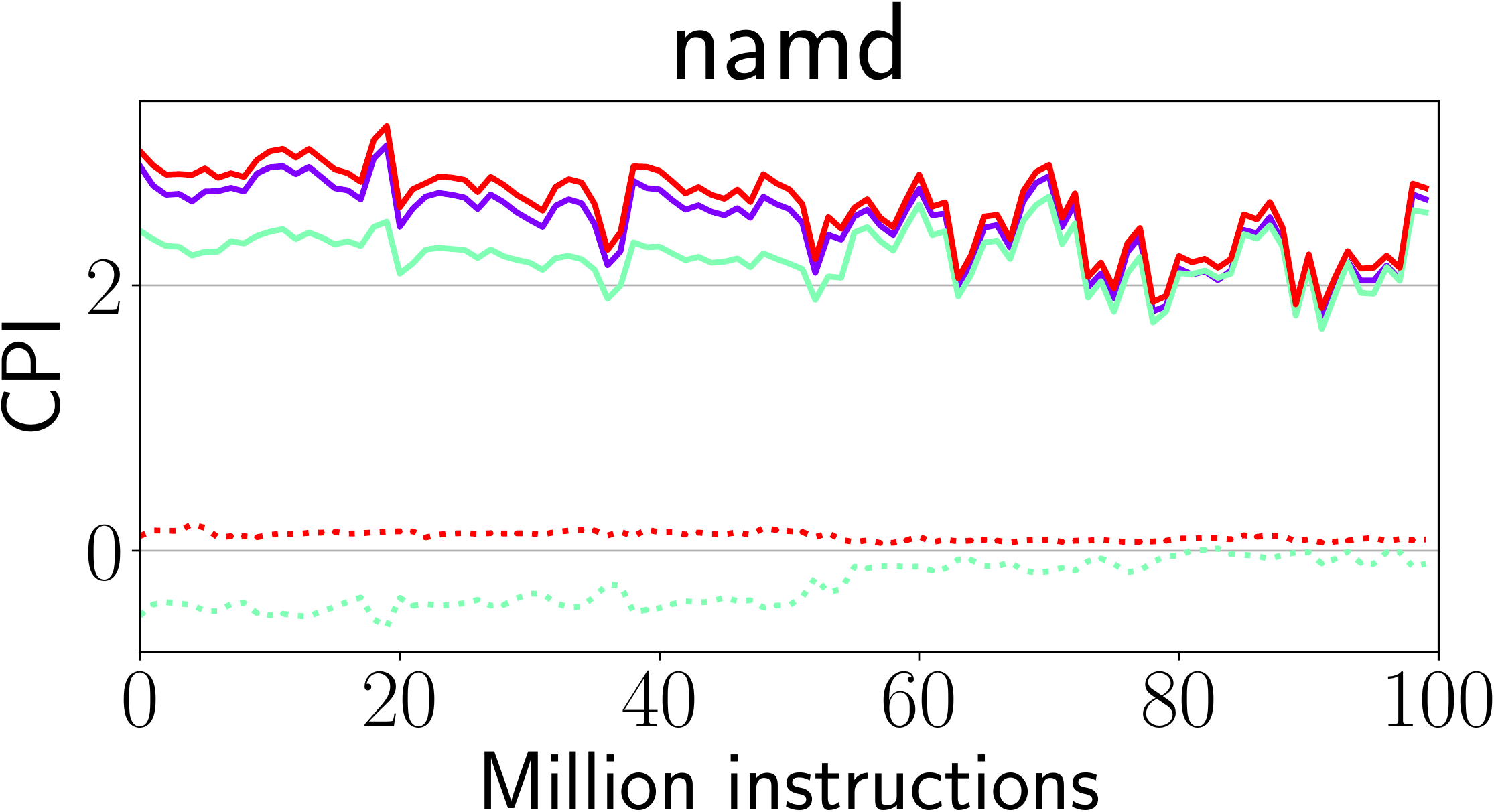}
  \hfil
  \includegraphics[width=0.2\textwidth]{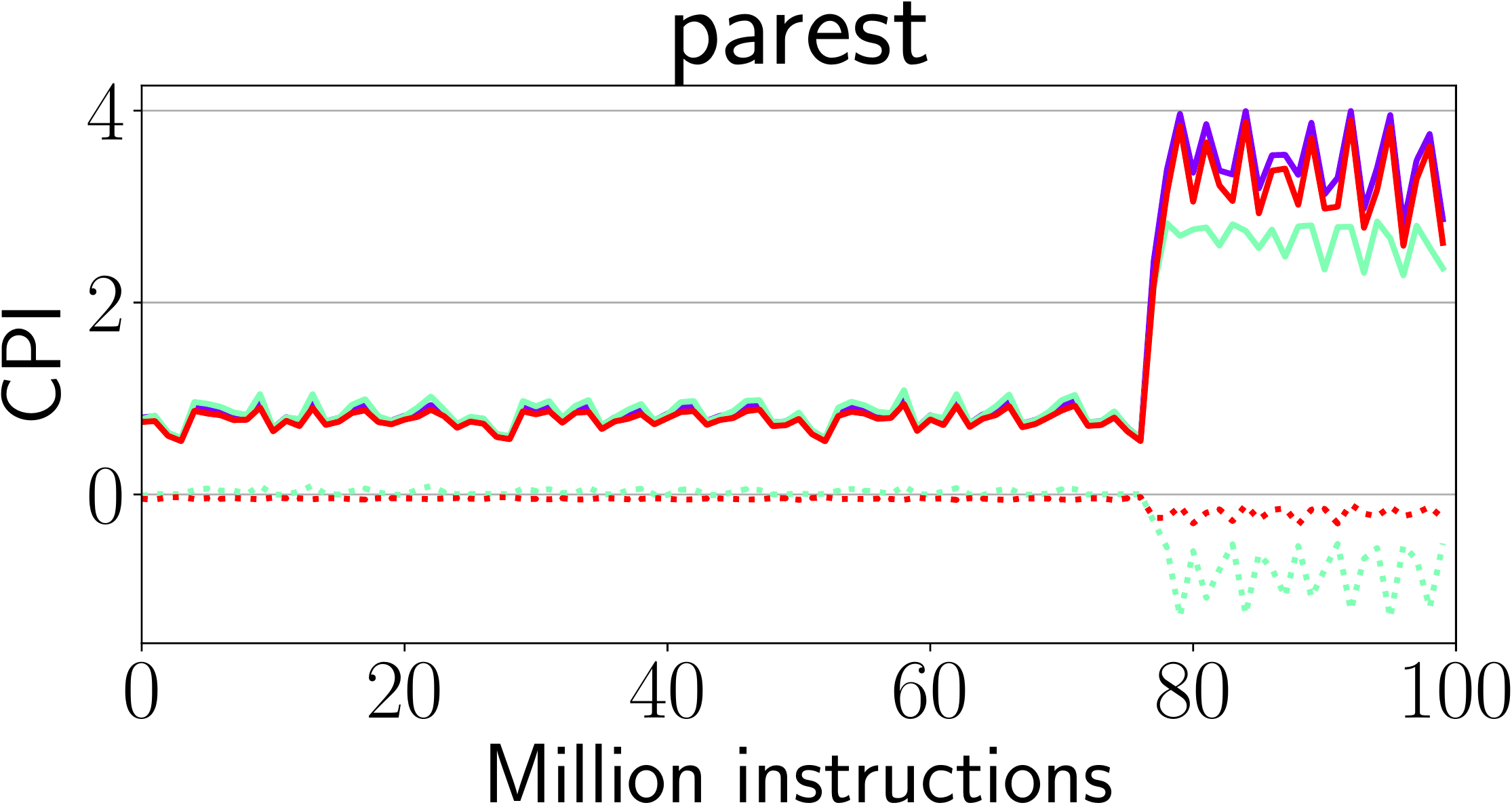}
  \hfil
  \includegraphics[width=0.2\textwidth]{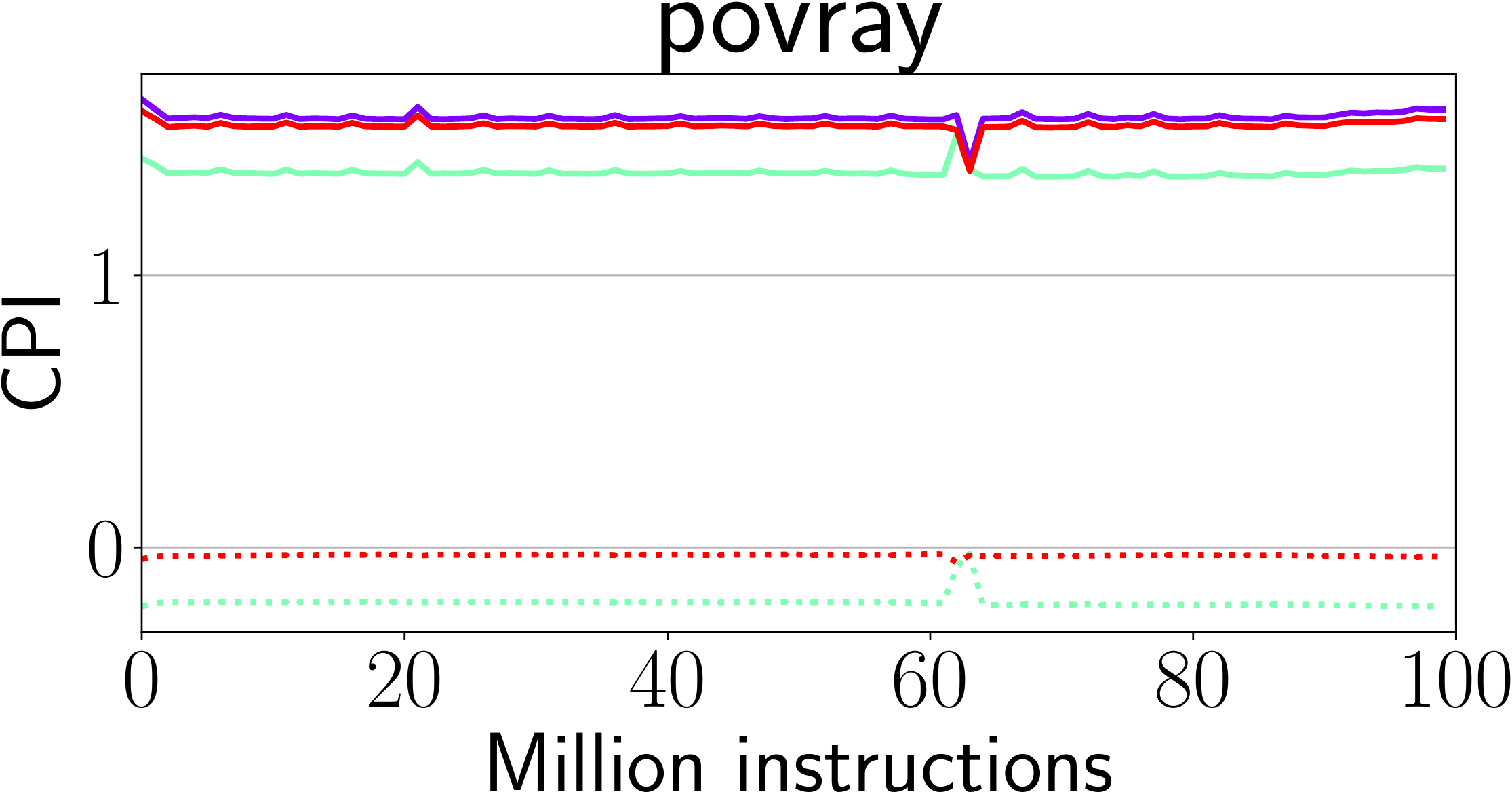}
  \hfil
  \includegraphics[width=0.2\textwidth]{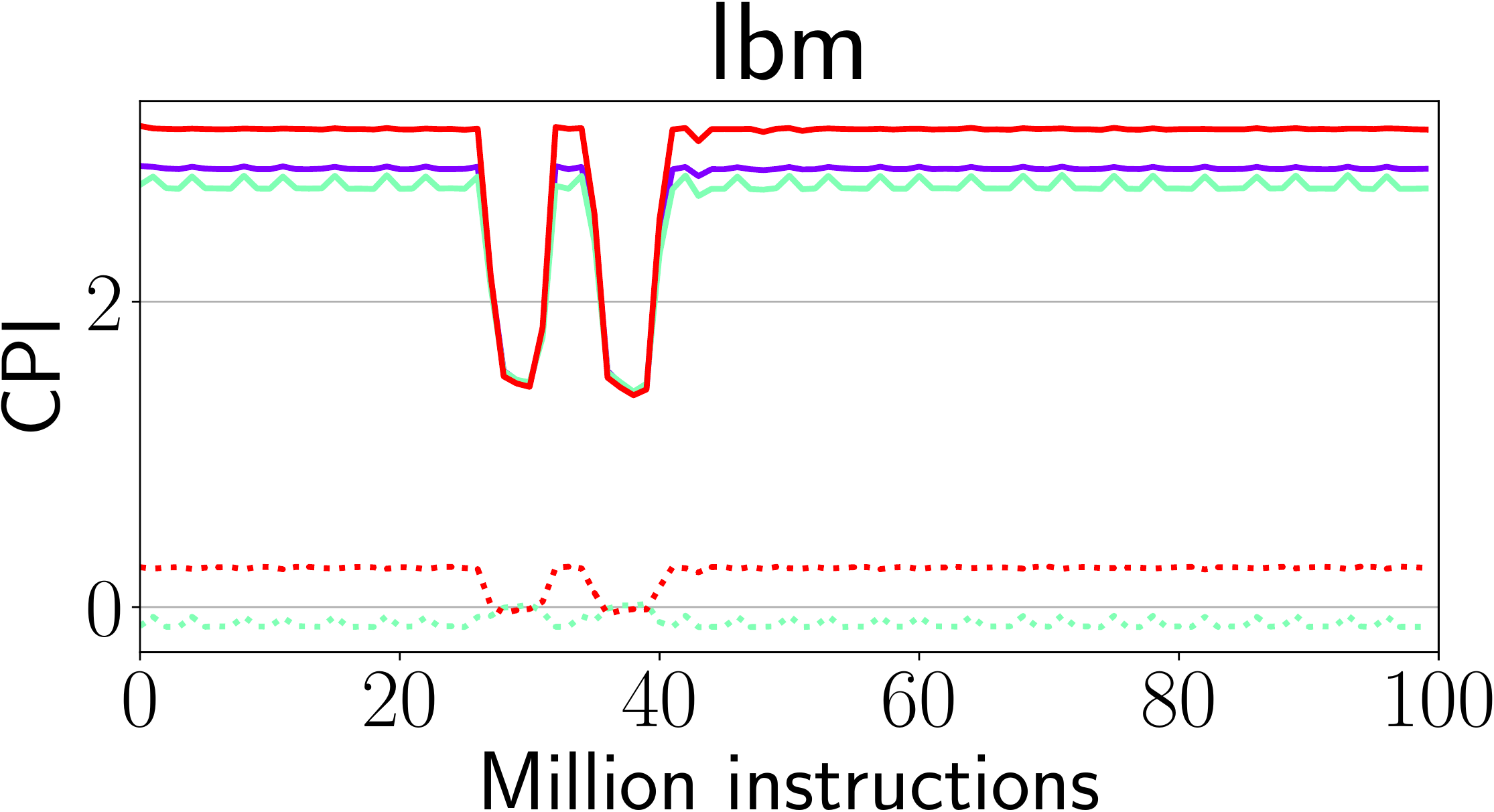}
  \hfil
  \includegraphics[width=0.2\textwidth]{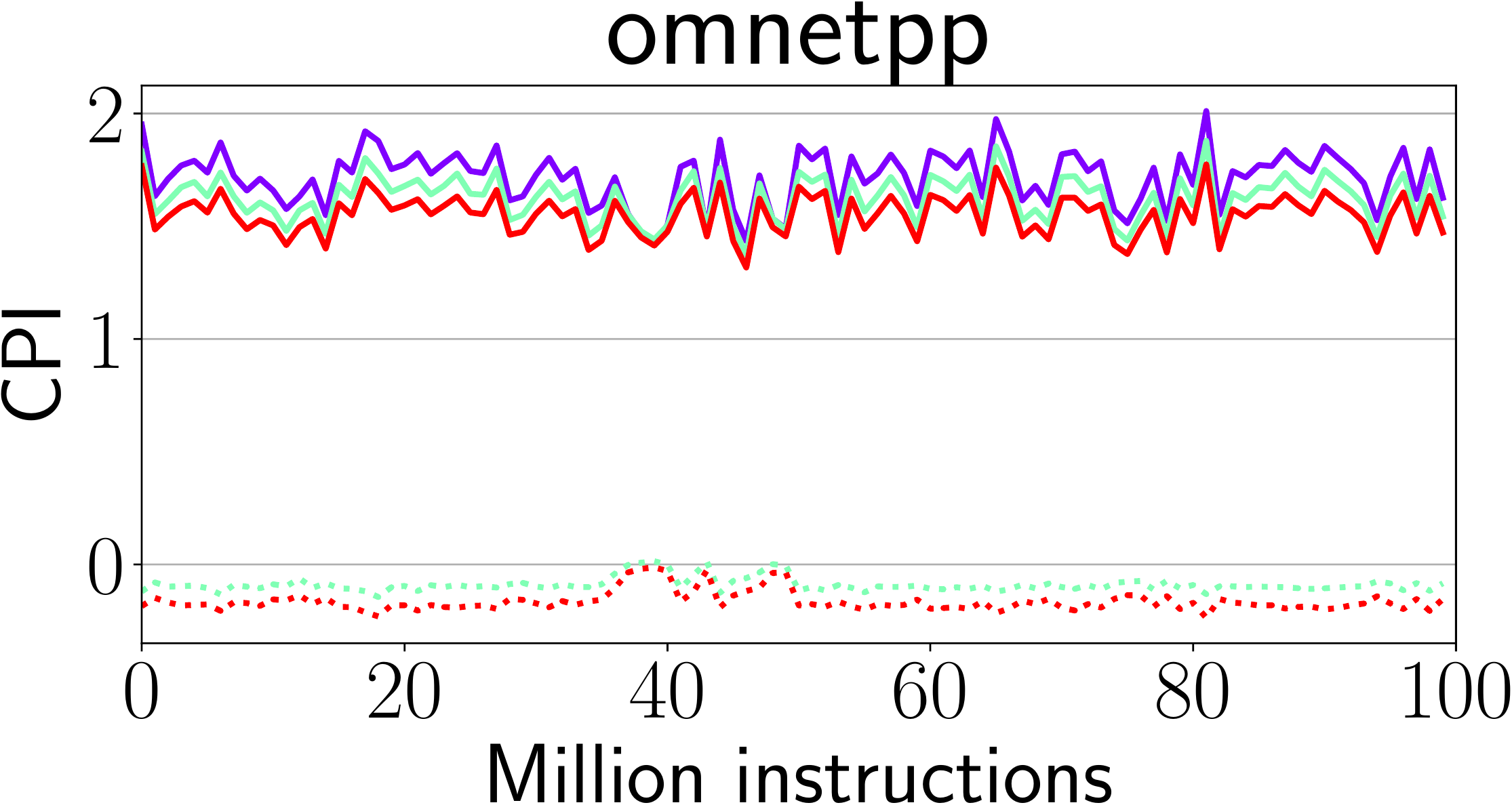}
}
\end{minipage}%

\begin{minipage}[t]{\textwidth}
\centerline{
  \includegraphics[width=0.2\textwidth]{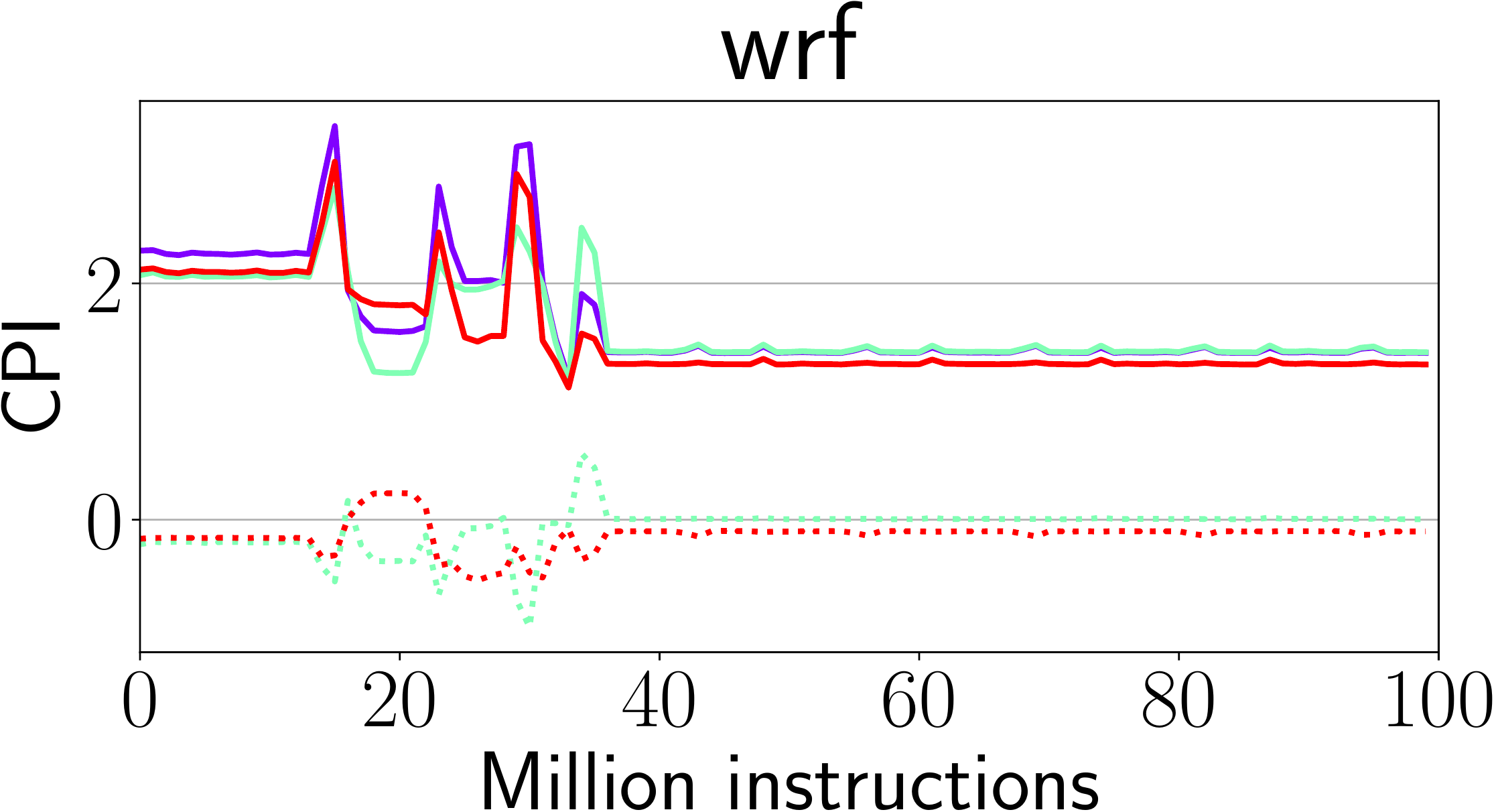}
  \hfil
  \includegraphics[width=0.2\textwidth]{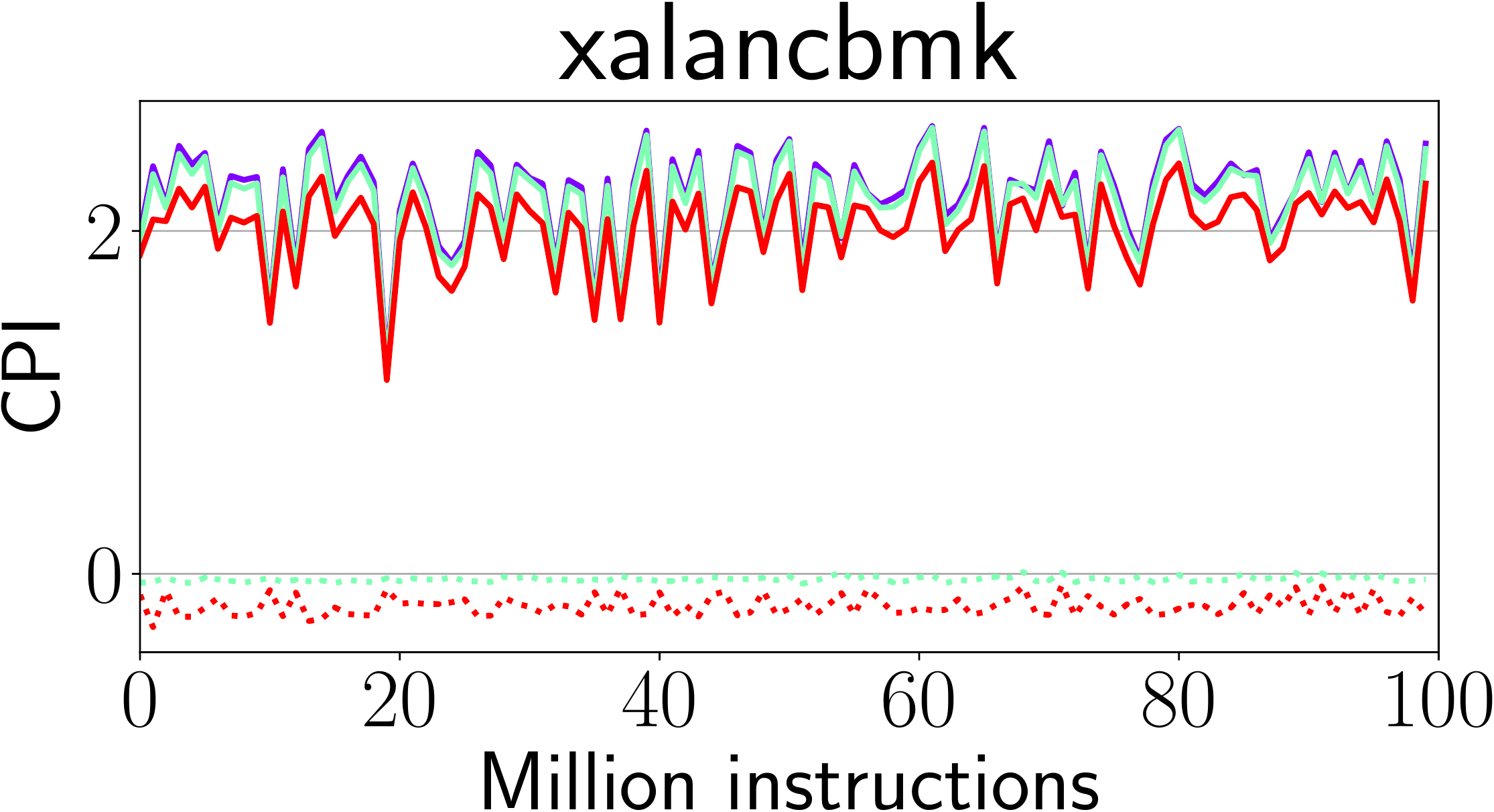}
  \hfil
  \includegraphics[width=0.2\textwidth]{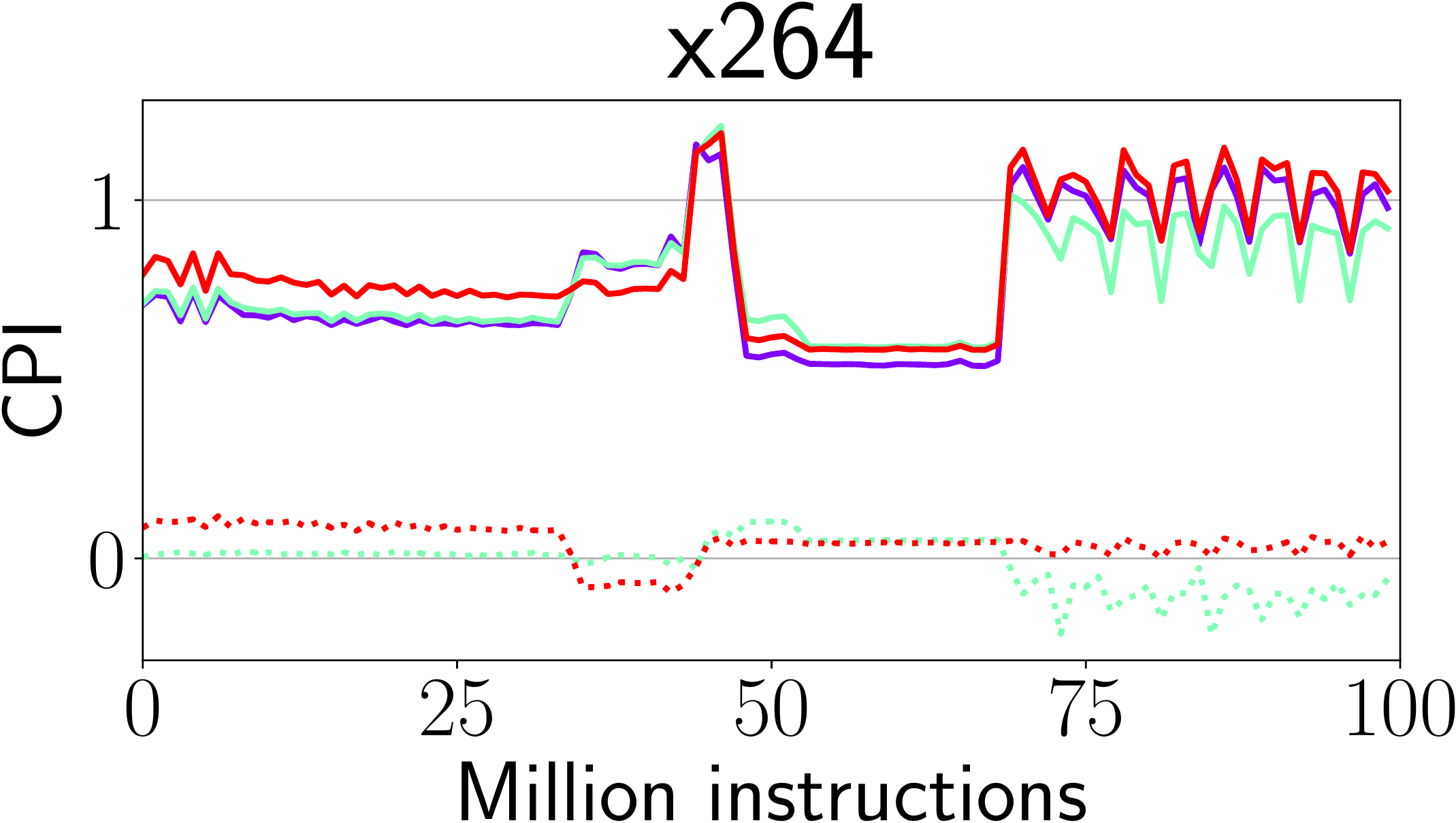}
  \hfil
  \includegraphics[width=0.2\textwidth]{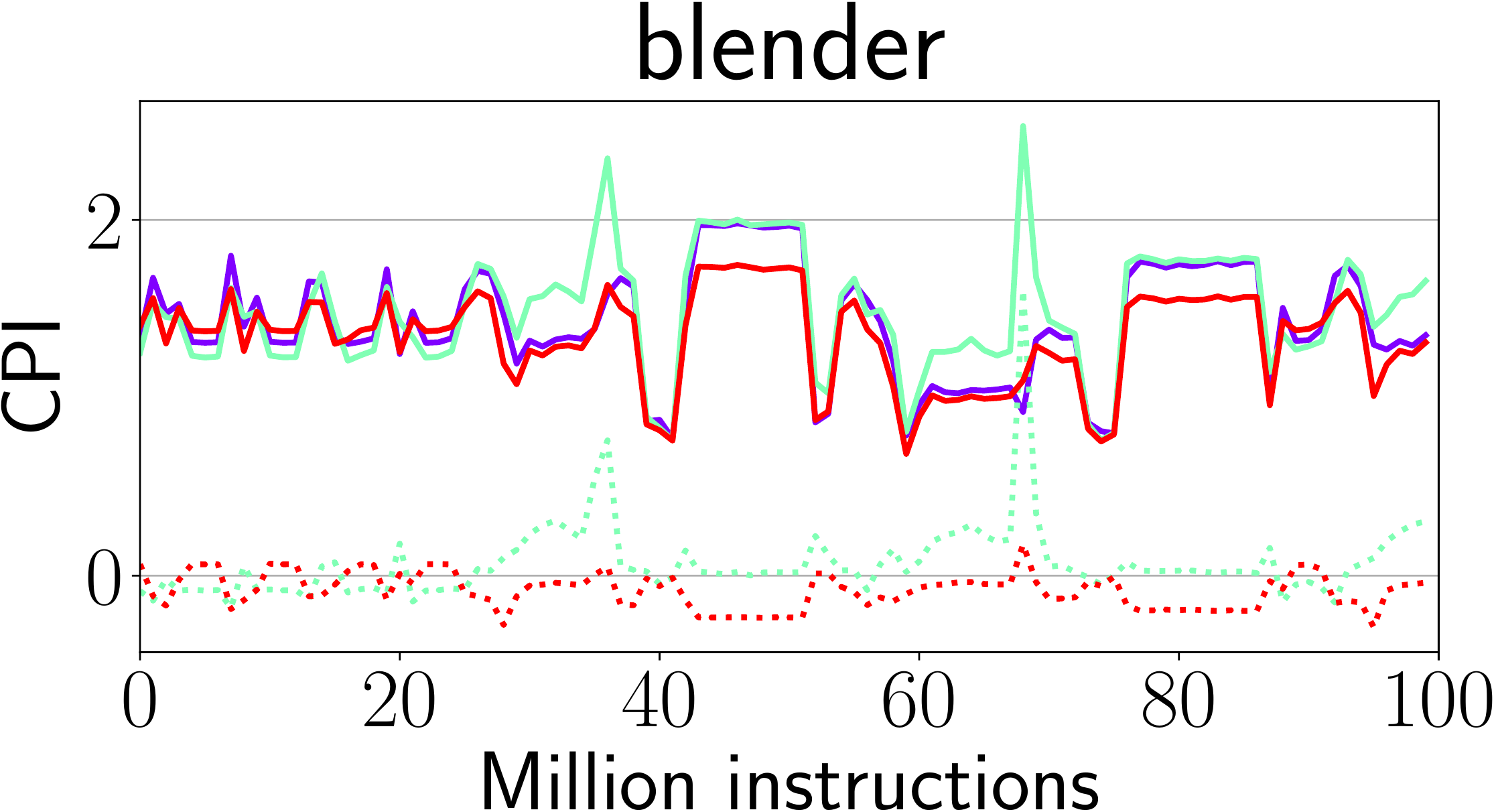}
  \hfil
  \includegraphics[width=0.2\textwidth]{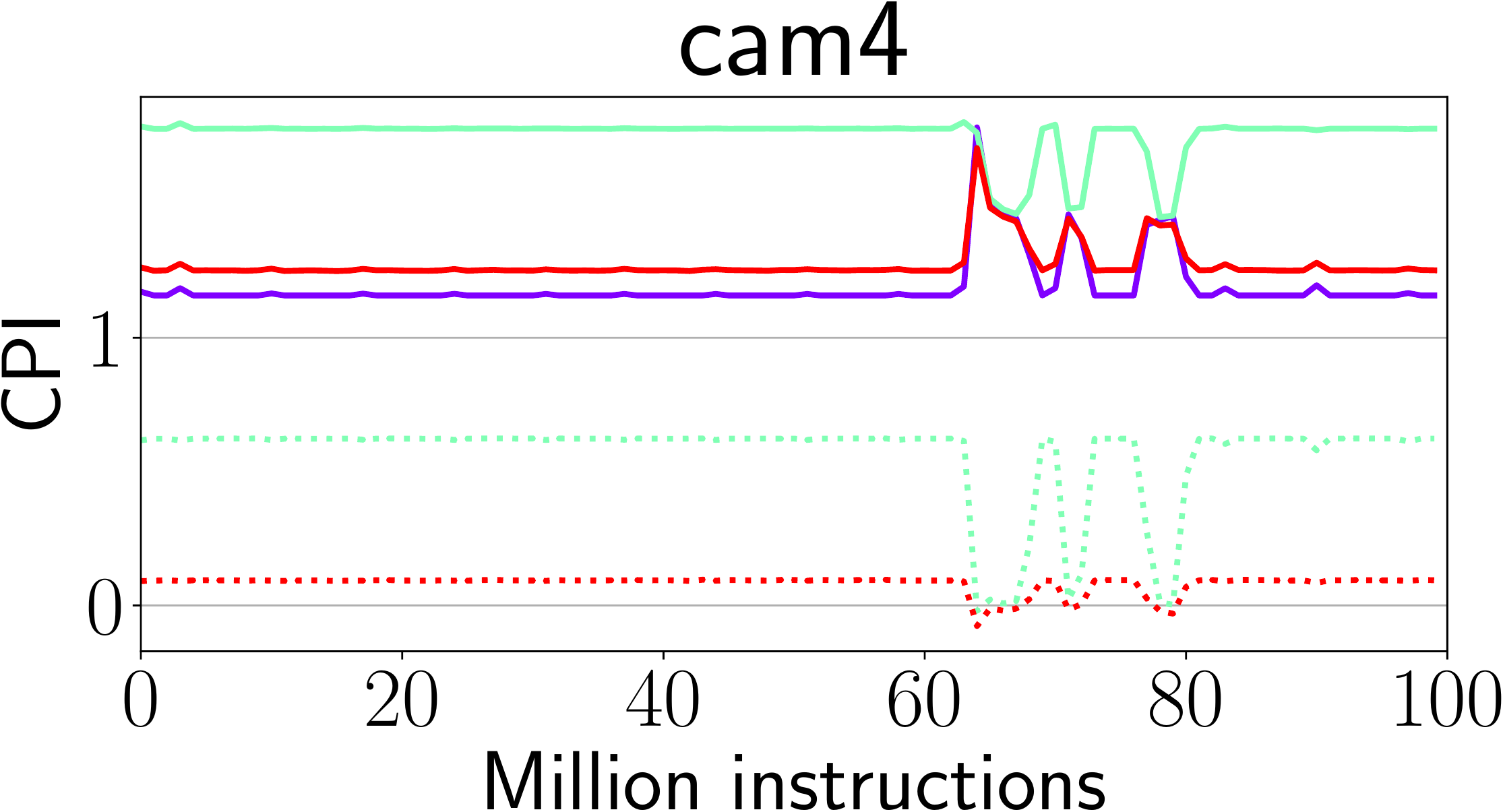}
}
\end{minipage}%

\begin{minipage}[t]{\textwidth}
\centerline{
  \includegraphics[width=0.2\textwidth]{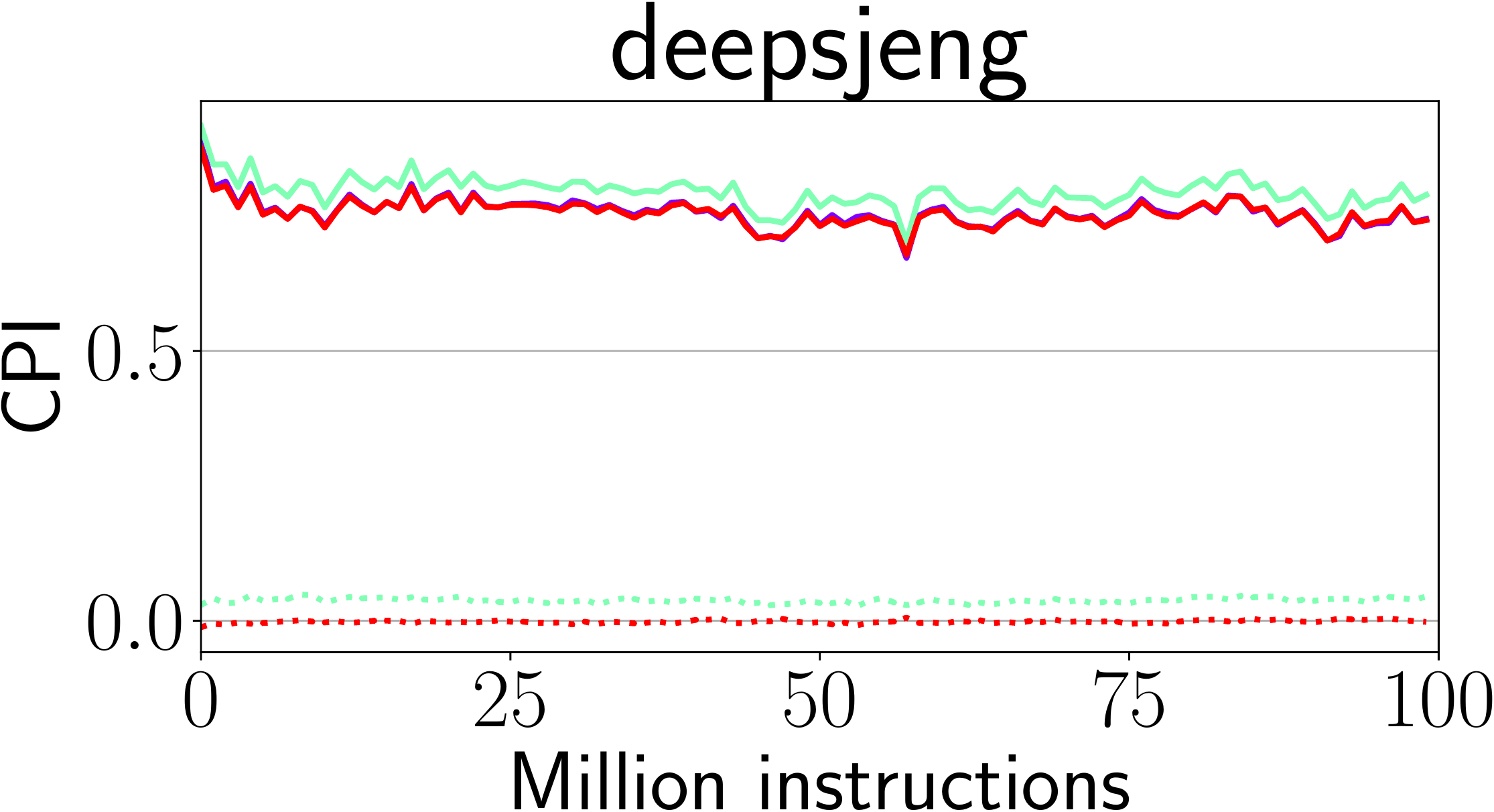}
  \hfil
  \includegraphics[width=0.2\textwidth]{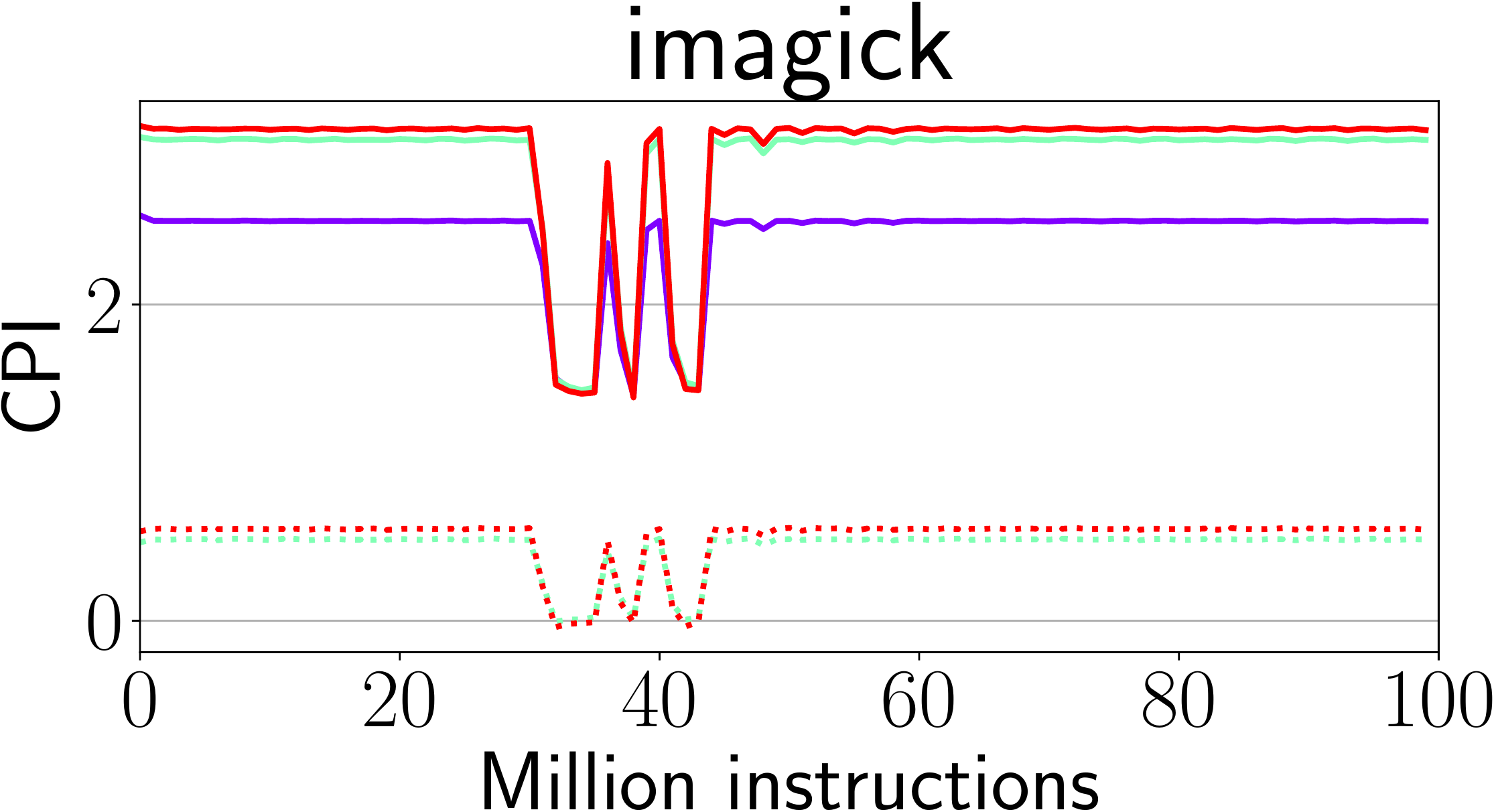}
  \hfil
  \includegraphics[width=0.2\textwidth]{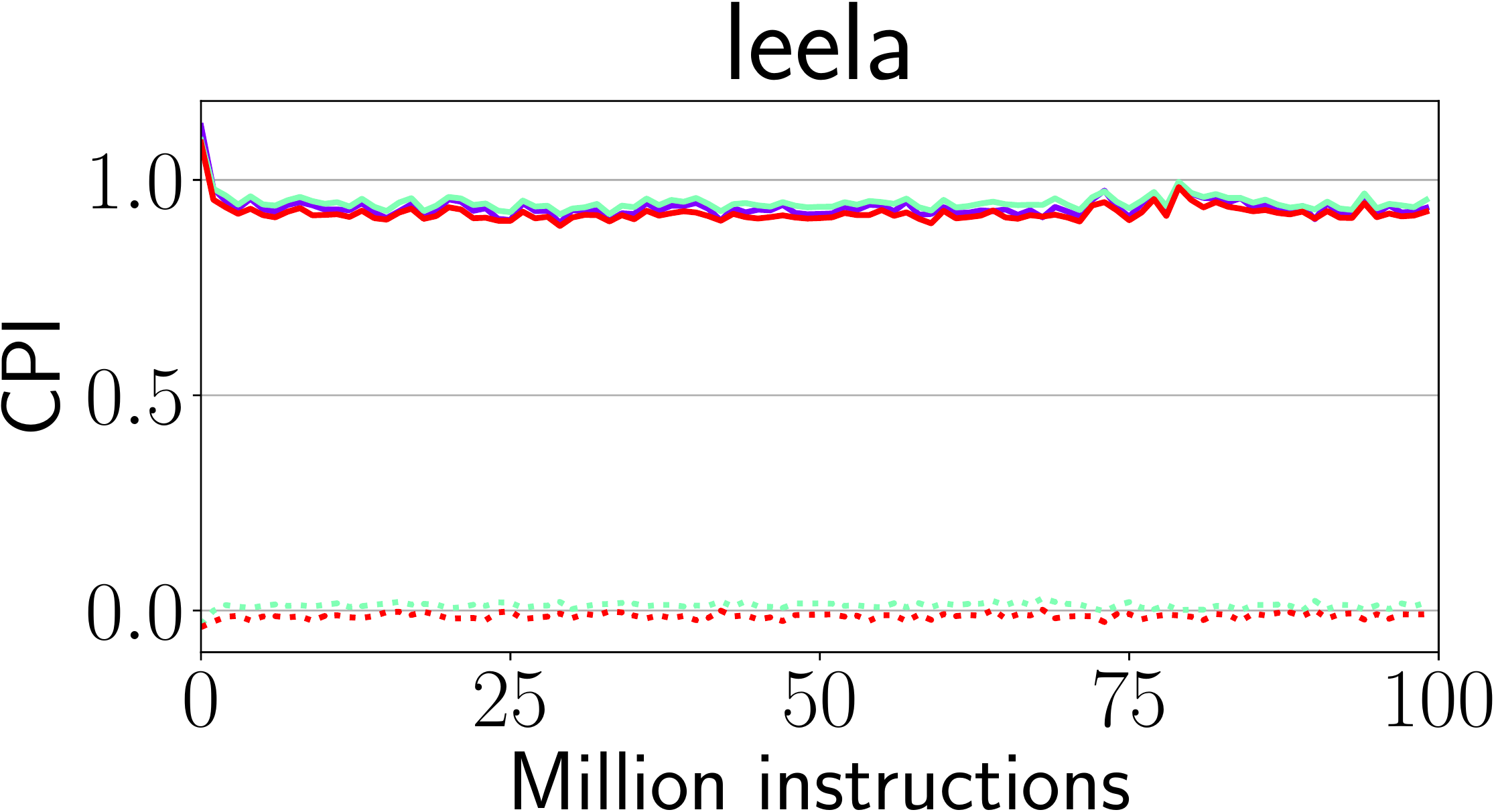}
  \hfil
  \includegraphics[width=0.2\textwidth]{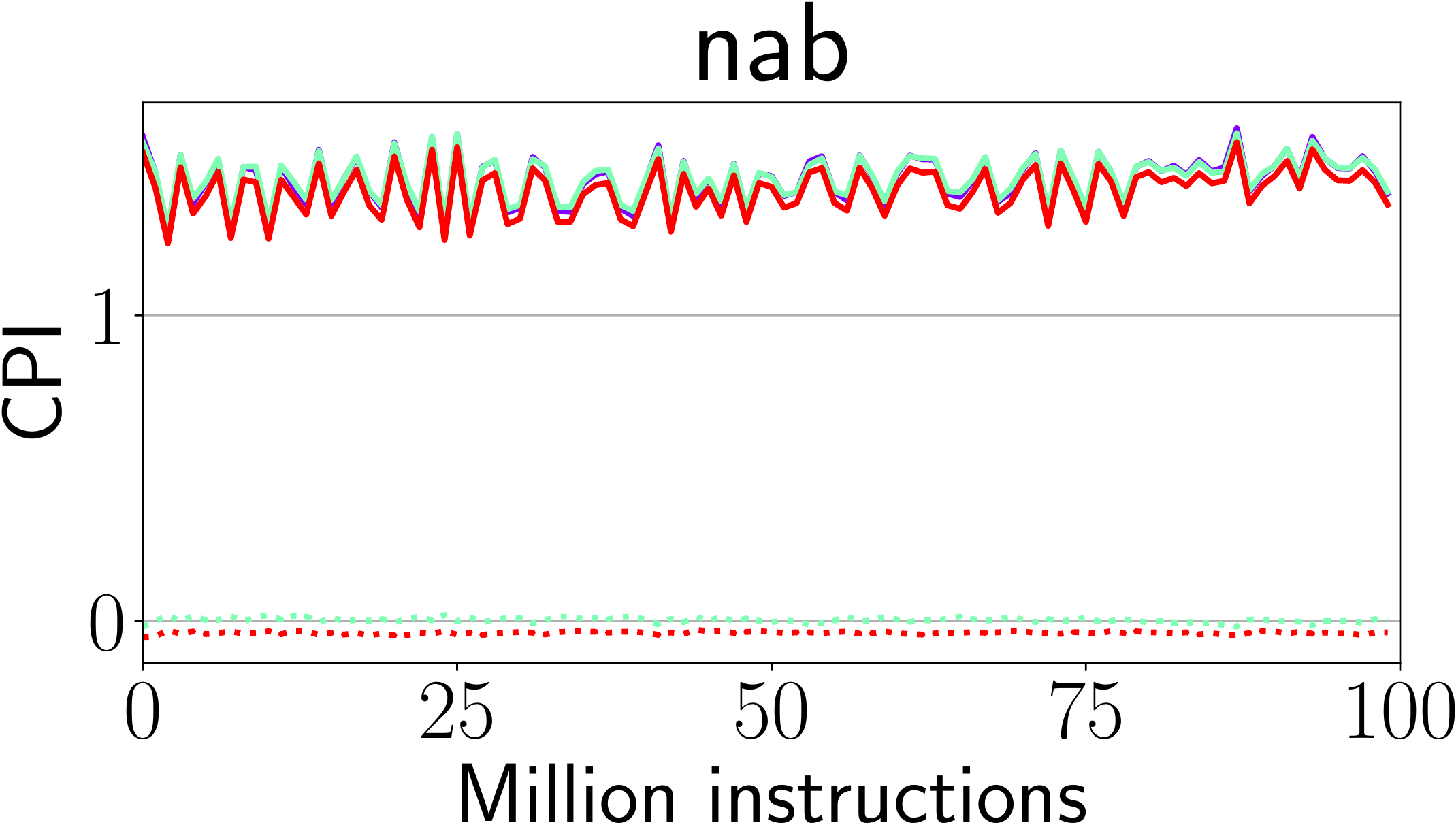}
  \hfil
  \includegraphics[width=0.2\textwidth]{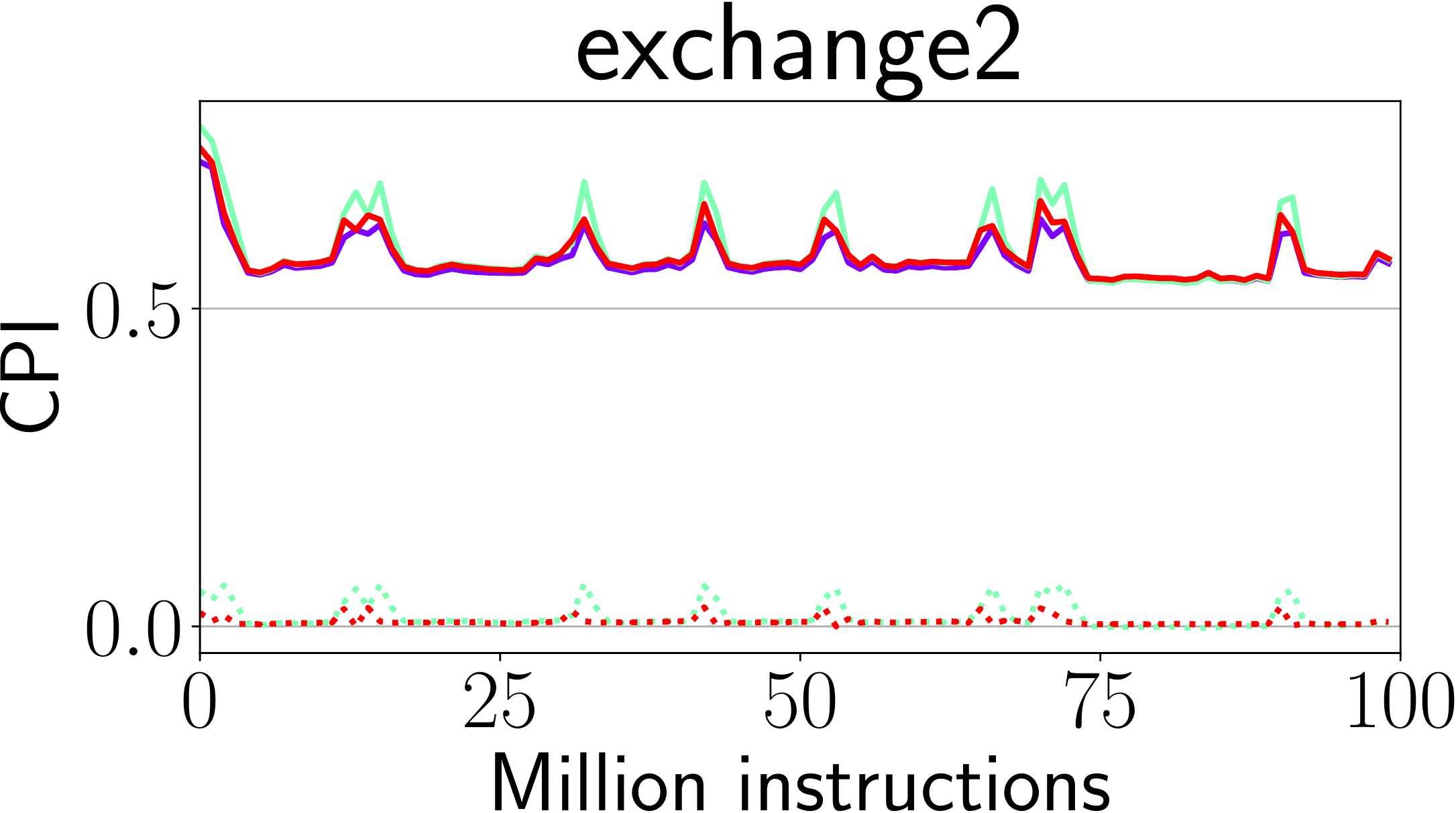}
}
\end{minipage}%

\begin{minipage}[t]{\textwidth}
\centerline{
  \includegraphics[width=0.2\textwidth]{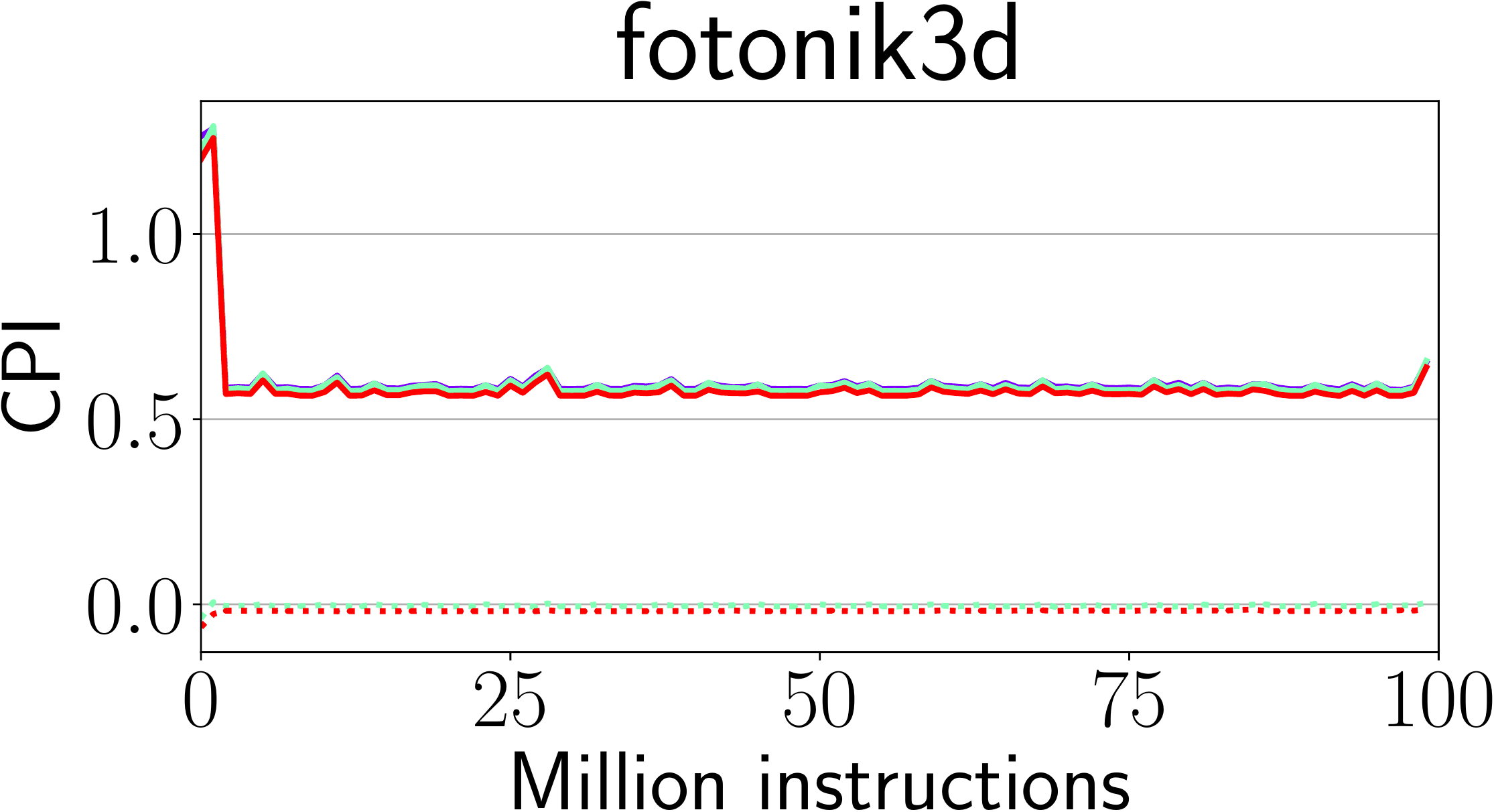}
  \hfil
  \includegraphics[width=0.2\textwidth]{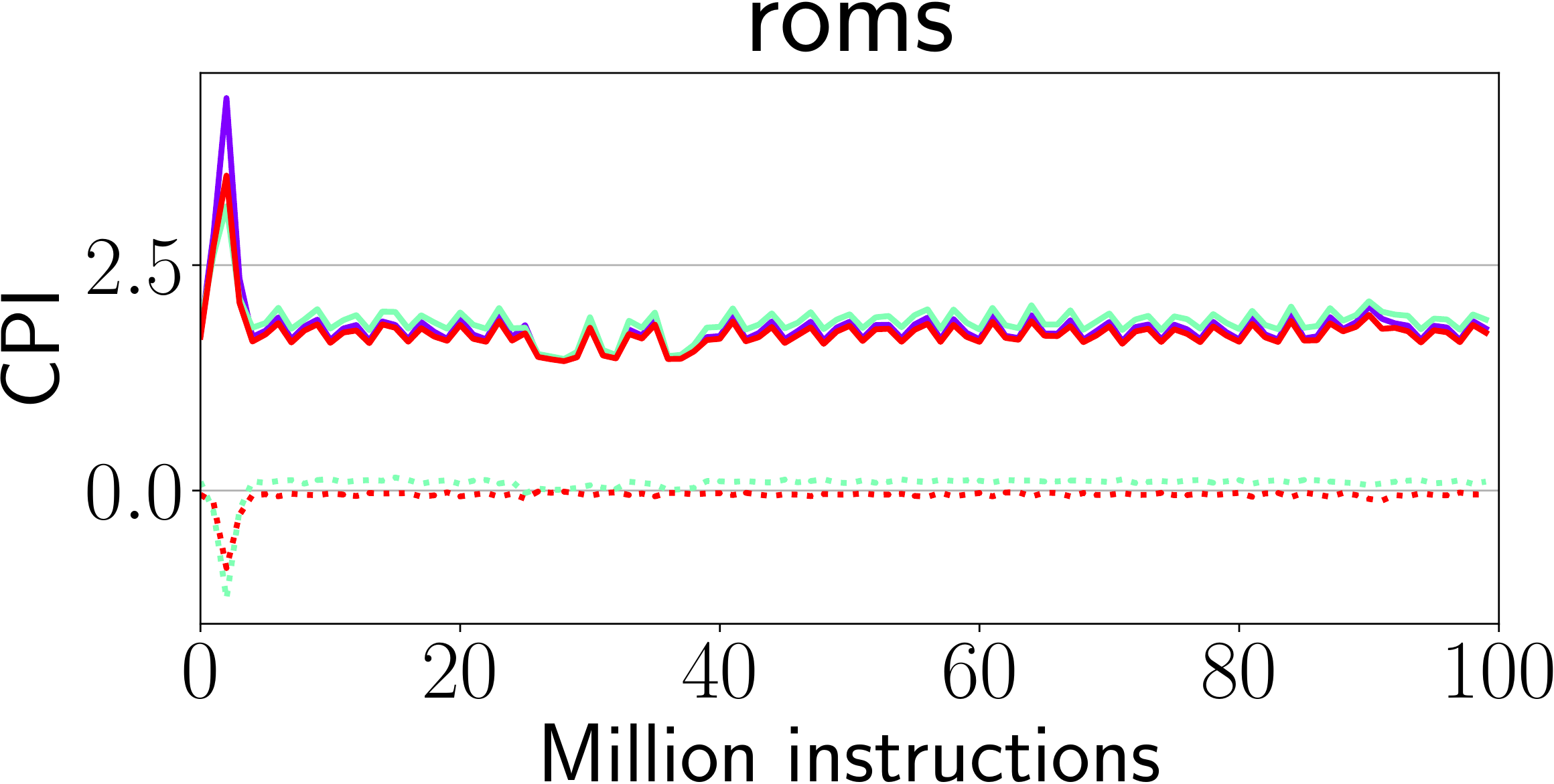}
  \hfil
  \includegraphics[width=0.2\textwidth]{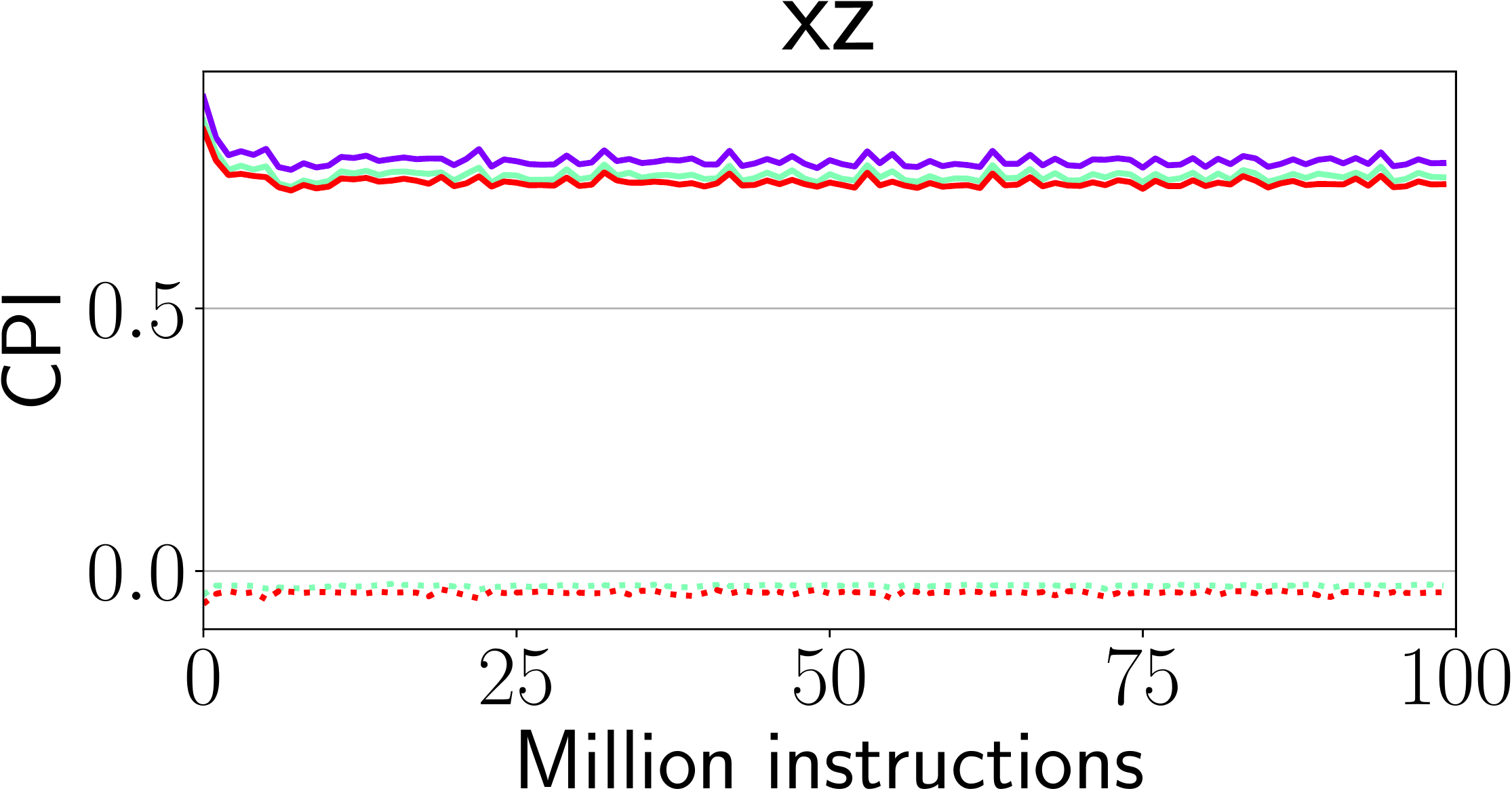}
  \hfil
  \includegraphics[width=0.2\textwidth]{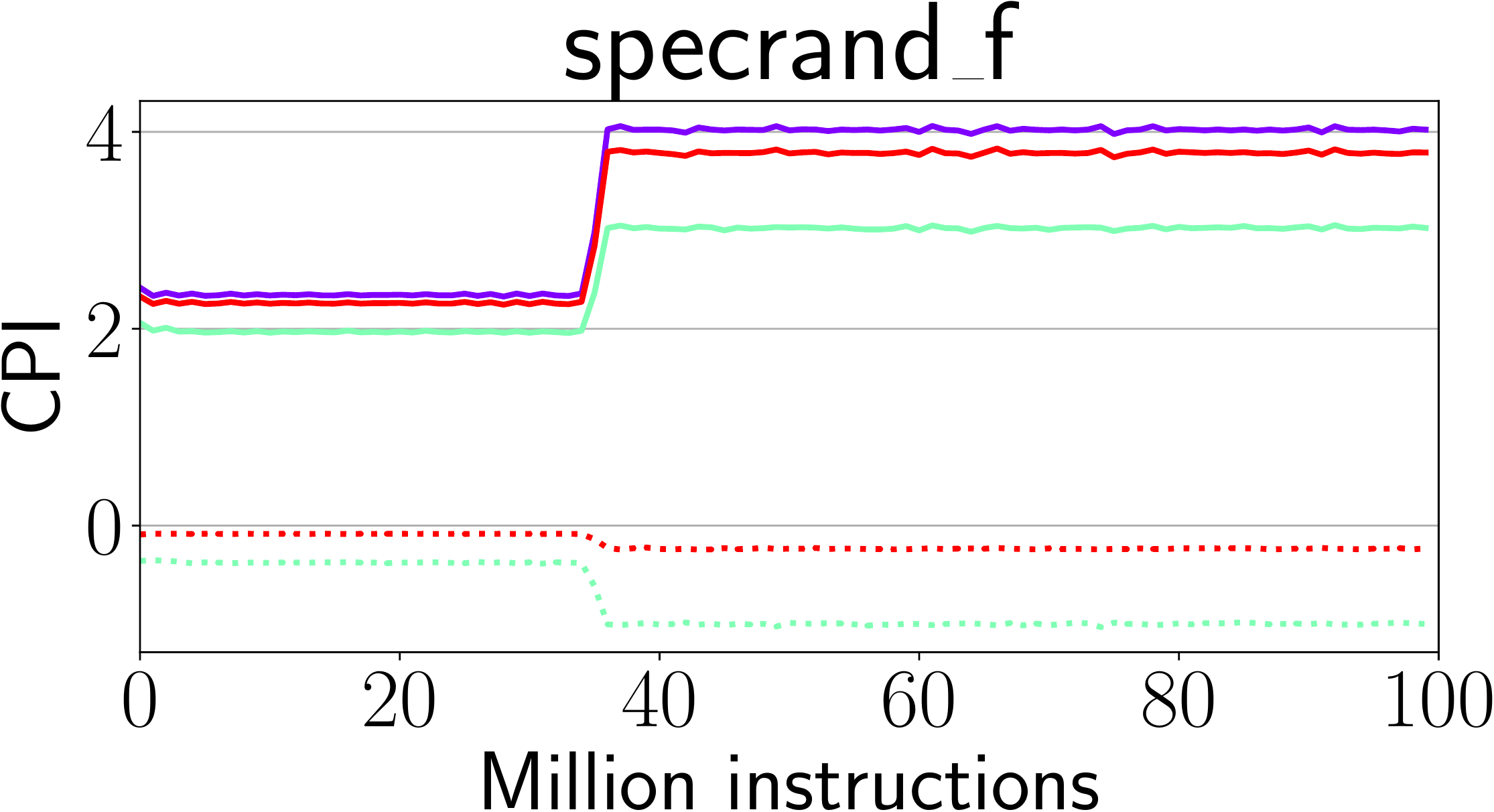}
  \hfil
  \includegraphics[width=0.2\textwidth]{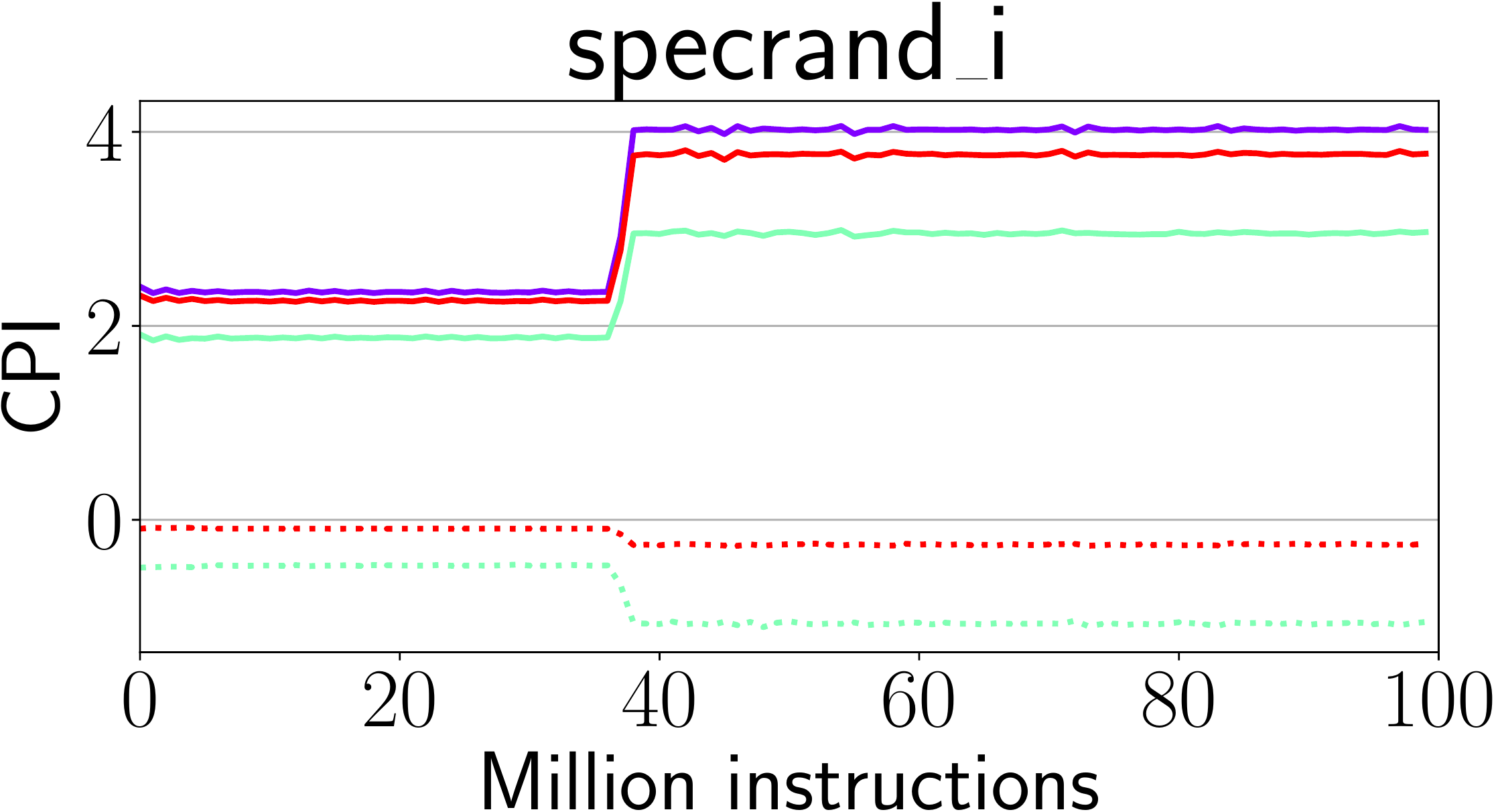}
}
\end{minipage}%

\caption{CPI variation during the simulation of 100 million instructions. The
solid lines show simulated CPI curves of gem5 and \simnet{} models. The dotted
lines show the simulation errors of \simnet{} models, calculated by subtracting
the CPIs of \simnet{} models by those of gem5.}
\label{fig:cpiphase}
\end{figure*}

\subhead{Phase Level Accuracy.}
To verify the simulation accuracy with respect to execution phases, Figure
\ref{fig:cpiphase} studies the CPI variation under C3 and RB7 models for all 25
benchmarks.
Particularly, we calculate the average CPI every 1 million instructions and
plot these CPIs over the total simulation length of 100 million instructions.
As observed in Figure \ref{fig:cpiphase}, benchmarks have either steady curves
(e.g., {\tt povray}, {\tt leela}), high CPI variations (e.g., {\tt perlbench},
{\tt gcc}), phased behaviors (e.g., {\tt bwaves}, {\tt specrand}), or mixes of
them.

For most benchmarks, we observe that \simnet{}'s CPI curves almost perfectly
match those of gem5, especially those using RB7 (i.e., red dotted lines are
always close to 0).
This phenomenon happens to many highly variable benchmarks such as {\tt
xalancbmk}, which demonstrates \simnet{}'s ability to capture small CPI
variations during simulation.
For {\tt cam4} where C3 has the largest simulation error (see Figure
\ref{fig:accuracy}), a consistent error persists across most simulation
periods, while RB7 still has a CPI curve that resembles that of gem5.
These results show that \simnet{} not only can predict the overall performance
well, but it also generates insights, such as identifying execution phases and
performance bottlenecks.

We also observe that a period of inaccurate simulation does not necessarily
affect the simulation accuracy of the time periods that follow.
For instance, C3's simulation errors reduce to 0 for 3 short periods when
gem5's CPIs increase for {\tt cam4}.
Another example is {\tt cactuBSSN}, where C3 fails to simulate it from 20M to
50M accurately, but has an almost identical CPI curve to that of gem5 after
50M.

This observation is counter-intuitive at first glance.
Because previous prediction results are used to construct the input of latter
predictions through the instruction context, it is reasonable to expect the
prediction errors will propagate through the simulation.
We discover two reasons behind it.
First, the processor pipeline is emptied every once in a while due to events
such as branch misprediciton during simulation.
Upon these events, there are no context instructions, and the predicted latency
does not rely on previous prediction results.
As a result, the latency is easier to predict by \simnet{} models and thus the
simulation accuracy gets calibrated on these events.
Second, a well-trained \simnet{} model can self correct its errors throughout
the simulation because such self-correction appears in the training data
generated from real processors' behaviors.
For example, assume one instruction $I$ takes longer than it should and
prevents the next instruction $I_n$ from entering the processor earlier.
When $I_n$ enters, the processor pipeline is emptier than it should be, which
results in faster execution of $I_n$.
In such a scenario, $I$ is executed slower, while $I_n$ is executed faster.
Together, the total execution time calibrates towards the right direction.
These reasons prevent the prediction error from propagating, and thus ensure
\simnet{}'s accuracy during long simulation.

\subhead{Accuracy Against Hardware.}
When there is an actual hardware that a simulator intends to simulate, the
simulator accuracy can be validated against the hardware.
For this purpose, we evaluate the accuracy of \simnet{} under the gem5 A64FX
configuration.
The gem5 simulaton of A64FX is verfied to have an average absolute error of
6.6\% against the real A64FX processor across a set of benchmarks
\cite{kodama2019evaluation}.
Since their simulated benchmarks do not include SPEC CPU 2017 benchmarks, we
cannot directly calculate the simulation accuracy of \simnet{} against the
A64FX processor.
Instead, we deduce the accuracy of \simnet{} against A64FX as follows.
Under a reasonable assumption that the normalized CPI follows a normal/Gaussian
distribution, we get the following distributions from \simnet{} results and
\cite{kodama2019evaluation}: $CPI_\text{\simnet} / CPI_\text{gem5} \sim
\mathcal{N}(1.062, 0.016^2)$, $CPI_\text{gem5} / CPI_\text{A64FX} \sim
\mathcal{N}(1.013, 0.078^2)$.
Their product, $CPI_\text{\simnet} / CPI_\text{A64FX}$, which represents the
accuracy of \simnet{} against A64FX, has a mean of 1.060 and a standard
variance of $0.016$ \cite{SASPWEB2011}.
The expected average absoluate simulation error is 6.0\% under this
distribution, similar to that of gem5.

To give more contexts, a simulator is usually considered to be accurate if
simulation errors are around 10\%.
For example, ZSim reports an average error of 9.7\% against an Intel Westmere
CPU \cite{ZSim}, and \cite{ISPASS14:gem5err} reports a 13\% error of gem5
against an ARM Cortex-A15 system.
Although we cannot directly validate the accuracy of \simnet{} against A64FX,
the deduced average absoluate simulation error is similar to that of gem5 that
it learns from.
We contend the low simulation error of \simnet{} is sufficient to gain
confidence about its simulation results.

\subhead{Relative Accuracy.}
While the simulation accuracy against real hardware is a useful metrics,
simulators are often applied in design space exploration where no corresponding
hardware exists for verification.
In these cases, computer architects care more about the ``relative'' simulation
accuracy, which measures how accurate simulation results reflect the
performance variance under certain architecture changes.
For instance, how much the performance will improve with doubled cache sizes.
Section \ref{sect:case} will demonstrate that \simnet{} achieves excellent
relative accuracies using several case studies.

%

\begin{figure*}
  \begin{minipage}[b]{0.32\textwidth}
    \centering
    \includegraphics[width=\linewidth]{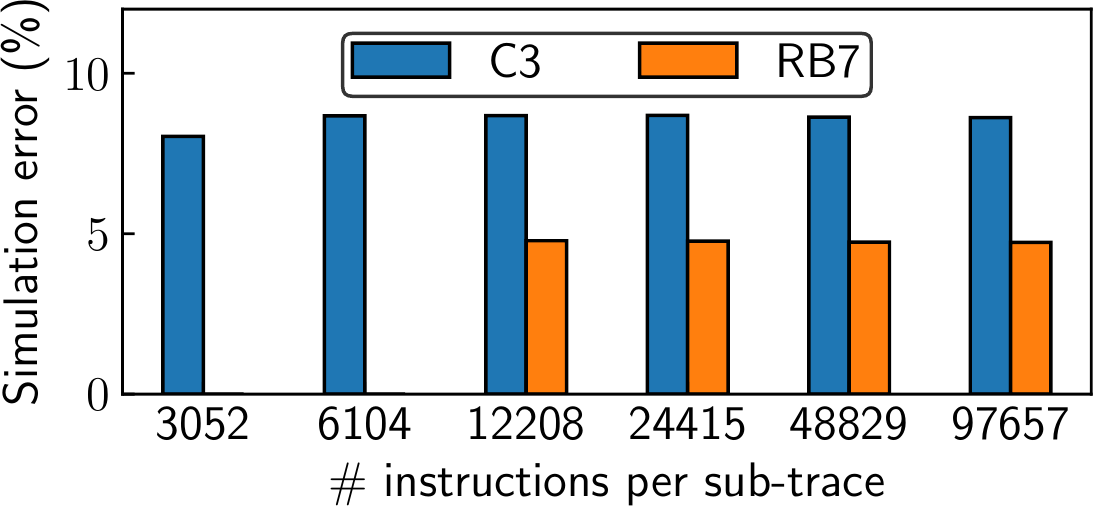}
    \caption{Average parallel simulation errors with various sub-trace sizes.}
    \label{fig:par_accuracy}
  \end{minipage}
  \hfill
  \begin{minipage}[b]{0.32\textwidth}
    \centering
    \includegraphics[width=\linewidth]{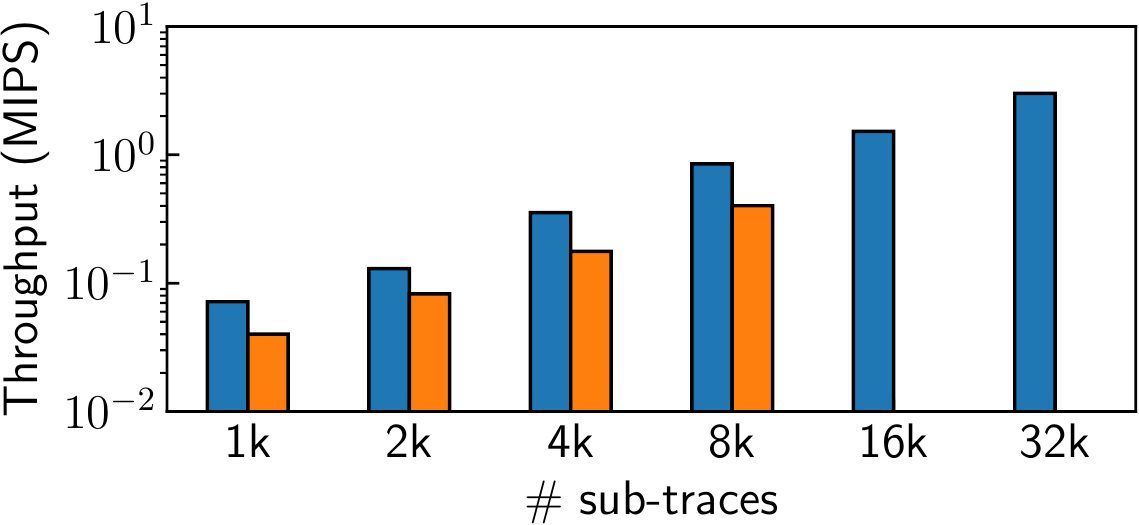}
    \caption{Simulation throughput with different sub-trace numbers.}
    \label{fig:subtracethroughput}
  \end{minipage}
  \hfill
  \begin{minipage}[b]{0.32\textwidth}
    \centering
    \includegraphics[width=\linewidth]{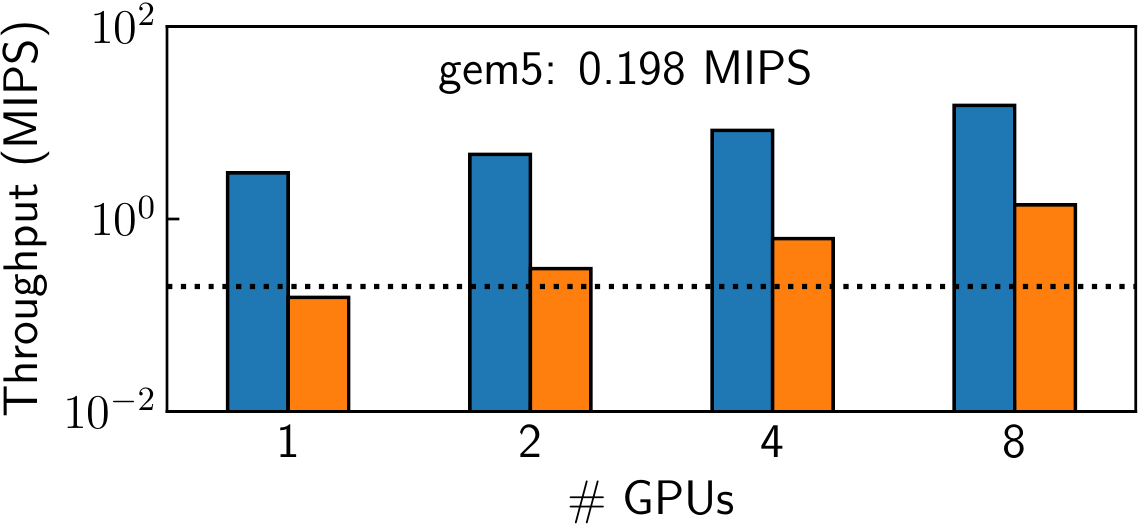}
    \caption{Simulation throughput with multiple GPUs.}
    \label{fig:gputhroughput}
  \end{minipage}
\end{figure*}

\subsecspace
\subsection{Parallel Simulation}
\label{sect:eval:par}
\subsecspace

\subhead{Accuracy.}
When simulating a single benchmark, because the parallel simulator partitions
the input trace into multiple sub-traces, there is simulation accuracy loss
across sub-trace boundaries.
Figure \ref{fig:par_accuracy} studies how the overall simulation accuracy
varies with the number of instructions per sub-trace.
RB7 cannot have sub-traces that are smaller than 12k instructions because the
GPU memory cannot accommodate too many sub-traces.
As the results show, sub-traces of 3k instructions are sufficient to achieve
parallel simulation errors that are similar to sequential ones.
The parallel simulation errors vary in a small range with sub-traces of
different sizes (around 8\% for C3 and 5\% for RB7), which demonstrates the
reliability of parallel \simnet{}.


\subhead{Throughput.}
We evaluate the simulation throughput of parallel \simnet{} in terms of million
instructions per second (MIPS).
Figure~\ref{fig:subtracethroughput} evaluates the average throughput across all
benchmarks with various numbers of sub-traces using the same models.
The x and y axes are on the logarithmic scale.
Limited by the GPU memory capacity, we cannot evaluate C3 beyond 32k sub-traces
or RB7 beyond 8k sub-traces.
We observe that the simulation throughputs improve almost linearly when
increasing the number of sub-traces, because more sub-traces allow \simnet{} to
utilize both CPU and GPU resources more efficiently until it saturates them.

Figure~\ref{fig:gputhroughput} assesses the throughput scalability of \simnet{}
with multiple GPUs, where the horizontal black-dotted line marks the gem5
simulation throughput.
As ML inferences take a significant portion of time in \simnet{}, using
multiple GPUs improves both the inference and simulation throughputs.
Again, \simnet{} achieves near-linear speedup with the number of GPUs.
With eight GPUs, it achieves 15.1 and 1.4 MIPS with the C3 and RB7 models.
Correspondingly, this represents $76.2\times$ and $7.4\times$ improvement over
gem5.
We can further scale \simnet{} to distributed GPU systems easily for higher
throughputs, because very limited communication is involved.

Note that we evaluate the throughput when simulating a single benchmark above.
In practical simulation scenarios, computer architects usually need to simulate
many benchmarks as well as different configurations.
Our design of \simnet{} can naturally simulate different benchmarks and
configurations in parallel, which provides even more opportunities to exploit
parallelism.

\subhead{Comparison with CPU-based Parallel Simulation.}
Previous CPU-based simulators can make use of multi-core CPUs to simulate
multiple programs/threads in parallel \cite{gem5, ZSim, sst}.
However, they cannot simulate a single program/thread in parallel and their
parallelism is limited by the number of cores (dozens on modern CPUs).
In comparison, GPU-based \simnet{} is able to simulate tens of thousands of
traces in parallel on one GPU as shown in Figure \ref{fig:subtracethroughput}.
These traces can come from a single or multiple programs/threads.
As discussed below, such massive parallel simulation of \simnet{} benefits not
only simulation performance, but also power efficiency.

\subhead{Power Efficiency.}
GPU-based \simnet{} can also achieve higher or similar simulation throughputs
given a certain power/energy budget compared with traditional CPU-based
simulators.
On our experimental platform, \simnet{} has a simulation power efficiency of
4.7 and 0.44 KIPS/watt for C3 and RB7, while that of gem5 is 0.88 KIPS/watt.
C3 is the most power efficient model while having acceptable simulation
accuracy.
While an A100 GPU has a TDP of 400 watts, we expect that \simnet's power
efficiency can be further improved using consumer grade GPUs such as NVIDIA
GeForce series or ASIC ML accelerators.

\subsecspace
\subsection{Overhead Discussion}
\label{sect:eval:overhead}
\subsecspace

\begin{figure*}
  \centering
  \includegraphics[width=0.4\linewidth]{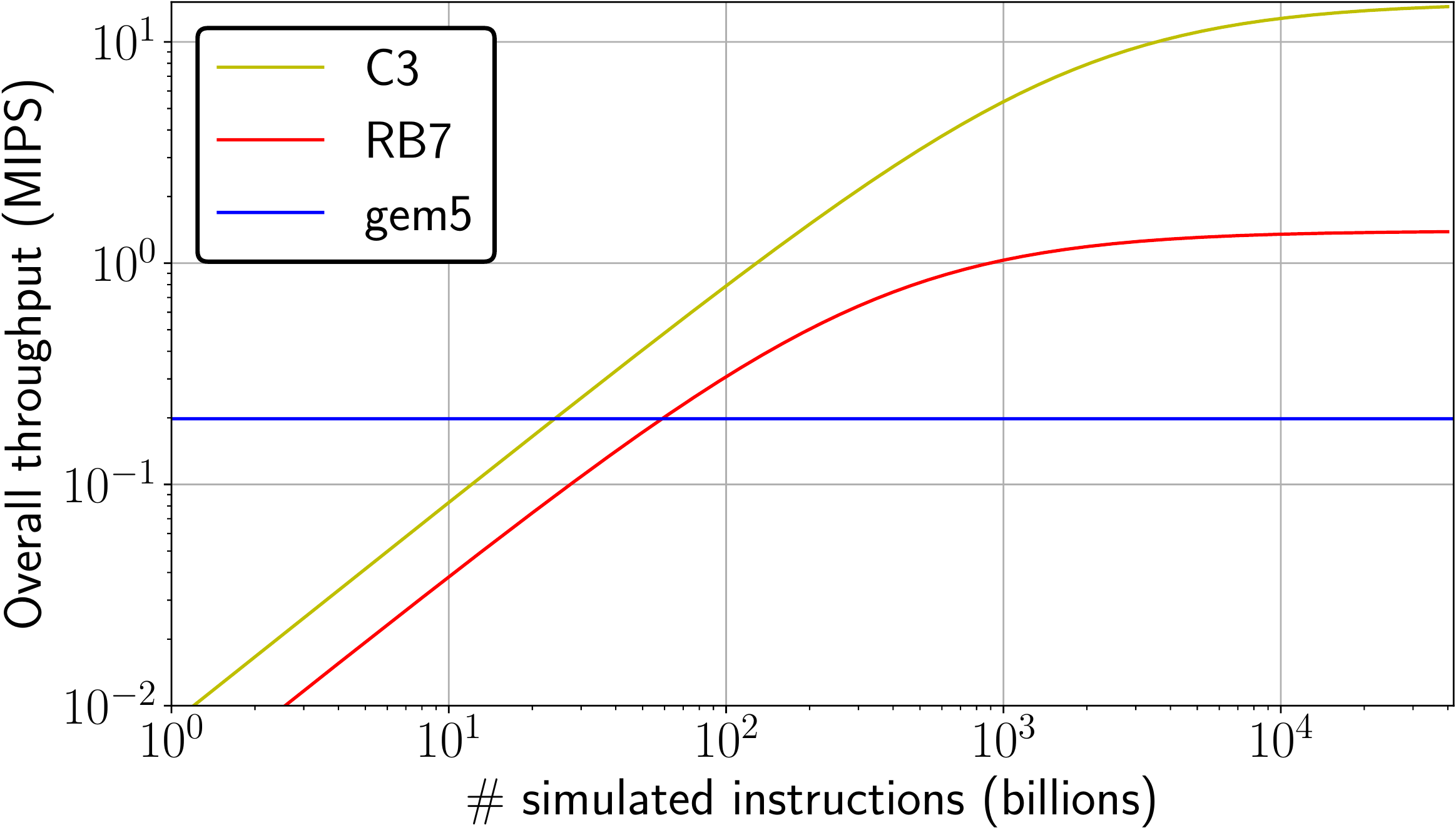}
  \caption{Overall simulation throughput under different instruction numbers.}
  \label{fig:train_overhead}
\end{figure*}

\subhead{Training Overhead.}
Figure \ref{fig:train_overhead} shows the overall throughputs of various
\simnet{} models that considers both simulation and training time.
It is calculated as $\frac{\text{\# simulated instructions}}{\text{training
time} + \text{simulation time}}$.
The training overhead amortizes with the increasing number of simulated
instructions.
The overall throughputs of \simnet{} exceed that of gem5 by 24 and 59 billion
instructions for C3 and RB7, and approach their ideal throughputs with zero
training overhead at trillions of instructions.
To put it into context, a typical SPEC CPU 2017 benchmark executes more than
one trillion instructions using the reference workload, and computer architects
typically need to simulate dozens of benchmarks under hundreds of
configurations, which means quadrillions of instructions.
Even with the help of statistical simulation tools such as SimPoint, simulating
trillions of instructions is still needed assuming the common practice of
simulating at least 100 million $\sim$ 1 billion instructions per benchmark.
The training overhead of \simnet{} is negligible in these use cases.
It is also worth noting that the training and simulation of \simnet{} can be
trivially scaled to large distributed systems, which will further reduce the
training overhead.

\subhead{Functional and History Context Simulation Overhead.}
Functional simulation can be accomplished using fast instruction set
simulators/emulators such as QEMU \cite{qemu}.
History context simulation is also fast because it only requires simplified
results, such as cache access levels, where simulating address tag comparison
and replacement is sufficient.
Our initial experiments and previous research show such simulation can be done
at $\sim100$ MIPS on a single CPU core \cite{DASC14:QEMU}, which is much larger
than \simnet{}'s simulation throughputs.
These overheads are therefore negligible.
Further acceleration of functional and history context simulation is possible
with GPUs, and we leave it for future works.

\subsecspace
\subsection{Impact of Features}
\label{sect:eval:feature}
\subsecspace

\begin{figure}[t]
    \centering
    \subfloat[To-be-predicted instruction.]{
        \includegraphics[width=0.34\linewidth]{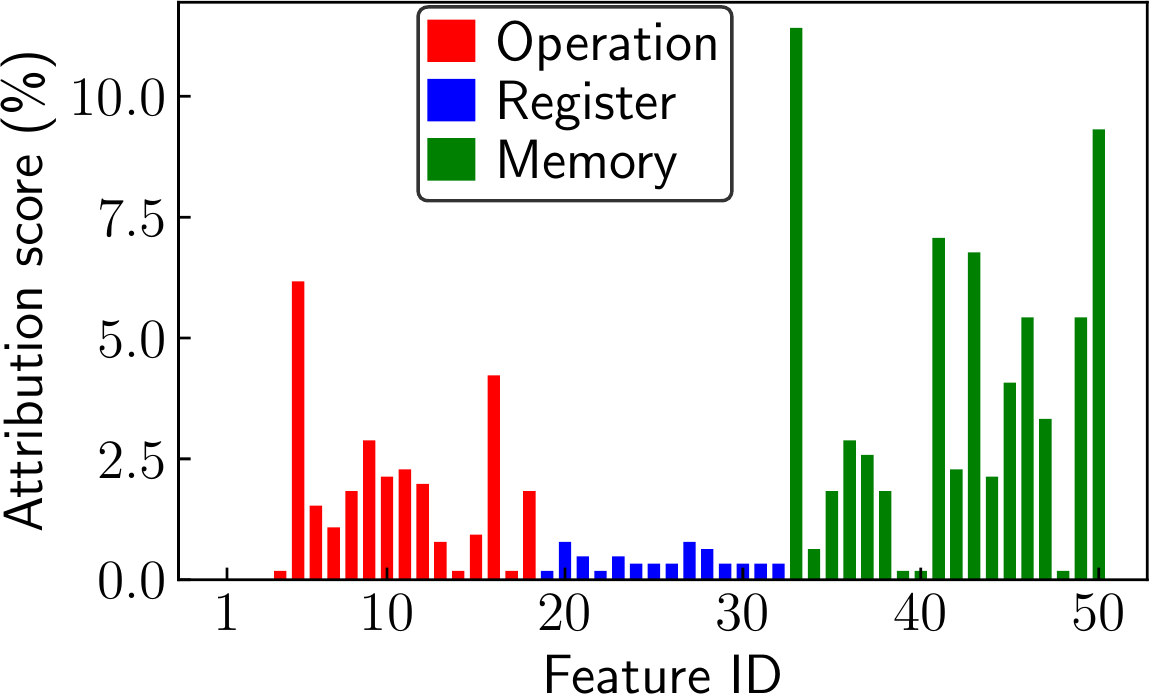}
        \label{fig:feature:cur}
    }
    \subfloat[Context instructions.]{
        \includegraphics[width=0.3\linewidth]{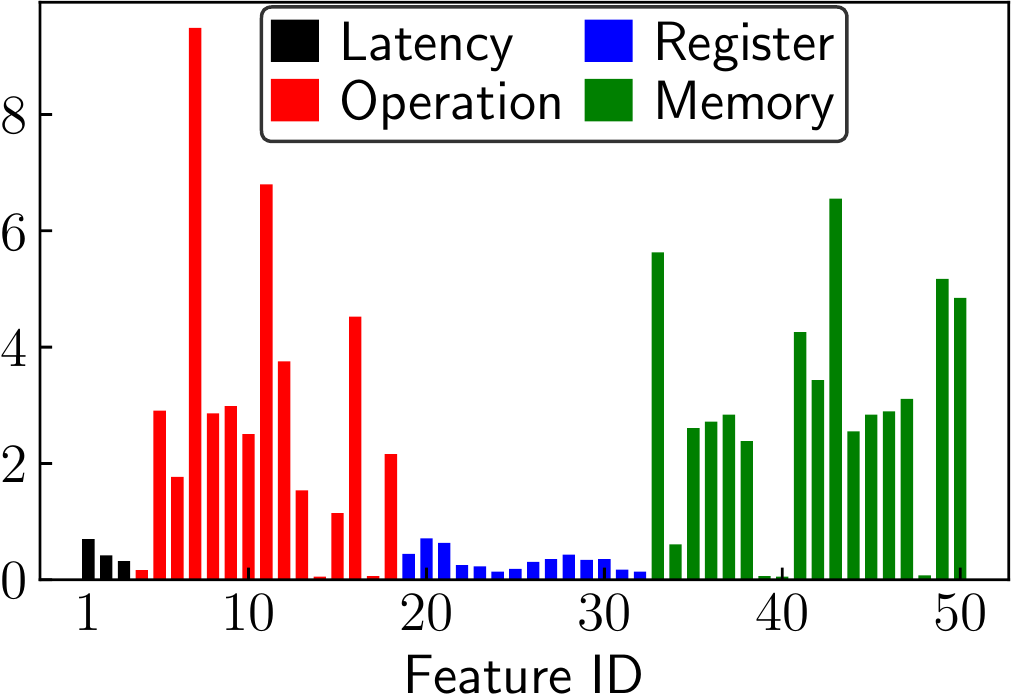}
        \label{fig:feature:cxt}
    }
    \caption{Feature attribution scores.}
    \label{fig:feature_importance}
\end{figure}

Figure \ref{fig:feature_importance} evaluates the contribution of each input
feature to the output for C3 using the SHapley Additive exPlanation (SHAP)
method \cite{NIPS2017_7062}.
SHAP's goal is to explain the prediction of an instance by computing the
contribution of each feature to the prediction.
It computes Shapley values~\cite{shapley1953value} using the coalitional game
theory.
Shapley value is the average marginal contribution of a feature value across
all possible coalitions.
We take the average of absolute Shapley values on training samples for each
feature to produce feature attribution scores.
Figure \ref{fig:feature:cur} and \ref{fig:feature:cxt} summarize the
attribution scores of to-be-predicted instructions and context instructions
separately.
We categorize the 50 features into latency, operation, register, and memory.
Memory and operation features generally have more impacts on the prediction
results.
The most influential feature of to-be-predicted instructions is the fetch
access level because the fetch latency depends on it.
For context instructions, the branch misprediction flag has the largest
attribution score as mispredicted branches need to flush the processor
pipeline.


\subsecspace
\subsection{Impact of Training Dataset Size}
\label{sect:eval:datasize}
\subsecspace

We also generate a large ML training dataset using 15 SPEC CPU 2017 benchmarks
instead of four.
Our results show that using the large dataset reduces the average simulation
error by 33\% at a cost of $3\times$ training time.
While larger training datasets further improve accuracy, we conclude that the
smaller dataset is enough to train accurate models and also requires less
training time.

The reason why a small training dataset is sufficient is because most
benchmarks use a variety of instructions which provide ample samples to train
the instruction latency predictor.
We expect reasonable ML prediction accuracy as long as there are adequate
samples to cover enough instruction and context scenarios, and the results show
that 4 benchmarks are enough to obtain sufficient scenarios.
Benchmark selection for the training set is also not critical.

\secspace
\section{Use Scenarios}
\label{sect:case}
\secspace

\simnet{} can be applied in many computer architecture research and engineering
scenarios.
First, many recent computer architecture efforts focus on caches or branch
predictors while other microarchitecture components are more sophisticated and
less likely to be subjects of change.
In such scenarios, pre-trained \simnet{} models can be directly applied as
caches and branch predictors are modeled in history context simulation, and no
additional training is required (i.e., the training overhead discussed in
Section \ref{sect:eval:overhead} does not exist.).
Second, when studying other microarchitecture parameters (e.g., ROB size) or
novel components, different parameter/configuration choices can be included in
the input of the model, so training a single model is sufficient to study all
variations.
We illustrate both use scenarios below, where the first two cases do not
require training, and the last case requires a one time training.

\begin{table}[t]
\scriptsize
\centering
\begin{tabular}{|l|c|c|c|c|c|}
    \hline
    & \multicolumn{3}{c|}{Simulated speedup} & \multicolumn{2}{c|}{Relative error range} \\
    \cline{2-6}
    & gem5 & C3 & RB7 & C3 & RB7 \\
    \hline
    BiMode\_l & 10.4\% & 11.2\% & 9.9\% & [-2.7\%, 3.7\%] & [-1.7\%, 1.4\%] \\
    \hline
    TAGE-SC-L & 12.3\% & 13.7\% & 12.4\% & [-4.0\%, 4.5\%] & [-0.7\%, 1.8\%] \\
    \hline
\end{tabular}
\caption{Simulated speedups of various branch predictors.}
\label{tbl:bp}
\end{table}


\subhead{Branch Predictor Study.}
We compare the simulated performance of two branch predictors using gem5 and
\simnet{}, including a large bi-mode branch predictor (BiMode\_l) and the
recently proposed TAGE-SC-L \cite{tage-sc-l}.
Their implementation in gem5 is used in \simnet{}'s history context simulation
to generate branch misprediction flags for ML models' input.
Table \ref{tbl:bp} shows the simulated average speedups across SPEC benchmarks,
where the speedup is calculated against the performance of a baseline bi-mode
branch predictor.
We observe that the average speedups obtained using \simnet{} are similar to
those using gem5.
Moreover, the right side of Table \ref{tbl:bp} shows the speedup error ranges
of individual benchmarks compared with gem5 results.
We observe that \simnet{} also predicts the speedups of individual benchmarks
well, especially under RB7.

\subhead{L2 Cache Size Exploration.}
We also simulate the performance impact of L2 cache sizes using gem5 and
\simnet{}.
Similar to the branch predictor case, \simnet{} accurately simulates the
relative speedup under cache sizes from 256 kB to 4 MB, and the average error
against gem5 is 0.8\%.

\subhead{ROB Size Exploration.}
In this experiment, the ML model input includes the ROB size as an additional
feature to account for its impact.
The training data are generated by running the same four SPEC benchmarks in
gem5 under various ROB sizes.
We train a C3 model to study the impact of ROB sizes.
Again, the simulation results of \simnet{} and gem5 agree with each other.
For example, the average performance improvement when increasing the number of
ROB entries from 40 to 80 and 120 is 1.2\% and 1.4\% under gem5.
Using \simnet{}, the corresponding speedups are 1.1\% and 1.5\%, which are very
similar.

\secspace
\section{Related Work}
\label{sect:relatedwork}
\secspace

\subhead{ML for Latency Prediction.}
Ithemal \cite{mendis2019ithemal} uses LSTM models to predict the execution
latency of static basic blocks.
The instructions within a block are fed into the model in the form of assembly,
such as words in NLP.
On top of Ithemal, DiffTune \cite{renda2020difftune} trains a differentiable ML
performance model to configure the simulator parameters to closely resemble a
target architecture.
These methods pose limits as they do not consider dynamic execution
behaviors, such as memory accesses and branches, which have significant impacts
on program performance.
They also target basic blocks with a limited number of instructions.
As a result, they are not applicable to a computer architecture simulator that
needs to simulate realistic processors and billions/trillions of instructions.

\subhead{ML for Application Performance Prediction.}
Ipek \etal propose using neural networks for application performance prediction
\cite{ASPLOS06:Ipek}.
Meanwhile, Lee \etal formulate nonlinear regression models for performance and
power prediction \cite{HPCA07:Lee, PPoPP07:Lee}.
Eyerman \etal propose inferring unknown parameters of mechanistic performance
models using regression, to balance between model accuracy and interpretability
\cite{ISPASS11:Eyerman}.
Mosmodel \cite{agbarya20-mosalloc} is a multi-input polynomial model used for
virtual memory research that can predict the program execution time given the
page table walking statistics.
Wu \etal use performance counters as the input of ML models to predict GPU
performance and power \cite{HPCA15:Wu}.
Nemirovsky \etal schedule threads based on ML-based performance models
\cite{SBAC-PAD17:Nemirovsky}.
Some approaches are proposed to predict a processor's performance/power based
on those obtained on different types of processors \cite{MICRO15:Ardalani,
SBAC-PAD14:Baldini, TECS17:Oneal} or with different ISAs
\cite{zheng2015learning, DAC16:Zheng}.

While these works build performance models on a per-program/input basis,
\simnet{} works at the instruction level.
Therefore, these application-centric approaches require generating training data and retraining models when target applications change, and the overhead of doing so is significant.
On the other hand, \simnet{} can directly simulate any
application, making it much more flexible.

\subhead{\bf ML for Other Architecture Research.}
In addition to the aforementioned uses, ML has been widely applied to
many other computer architecture aspects, including microarchitecture design and
energy/power optimization.
These applications are summarized in \cite{mlarchsurvey, wu2021survey}.

\subhead{Simulation with Statistical Sampling.}
Instead of simulating the entire program, statistical simulation selectively
simulates representative sampling units and infers the overall performance
from these sample simulation results statistically \cite{ICCD96:Conte}.
SMARTS \cite{ISCA03:SMARTS, Micro06:SimFlex} periodically switches between
detailed and functional simulation to obtain an accurate CPI estimation with
minimal detailed simulation.

SimPoint records the basic block execution frequencies of individual sampling
units and those of the whole program to select representative ones with the
aim that the selected samples capture the overall execution behaviors well \cite{ASPLOS02:SimPoint, SIGMETRICS03:SimPoint}.
Similarly, PinPoints uses dynamic binary instrumentation to find
representative samples for X86 programs \cite{MICRO04:PinPoint}, and
BarrierPoint applies sampling to multi-threaded simulation
\cite{ISPASS14:BarrierPoint}.
These methods require pre-analyzing the simulated program with a certain
input, while our ML-based simulator can be applied directly to any program and
input combination because of its instruction-centric approach.

One key challenge in statistical simulation is to keep track of the
microarchitecture state between detailed simulation fractions, especially cache
states.
To simulate the cache behavior accurately in statistical simulation, Nikoleris
\etal propose using Linux KVM to monitor the reuse distance of selected cache
lines \cite{MICRO19:CacheWarm}.
Similarly, Sandberg \etal leverage hardware virtualization to fast-forward
between samples, so different samples can be simulated in parallel
\cite{sandberg2015}.
These statistical simulation approaches can be used together with \simnet{} to
further accelerate the detailed simulation portions.
As an example, Section \ref{sect:eval} uses SimPoint and \simnet{} together.

\subhead{Traditional Simulation Acceleration.}
ZSim is an X86 simulator that supports many-core system simulation \cite{ZSim}.
It decouples the simulation of individual cores and resources shared across
cores, as well as adopts a simplified core model. As a result, it achieves
$\sim10$ MIPS for single-thread workload simulation on an Intel Sandy Bridge
16-core processor.
SST \cite{sst} distributes the simulation of different components across
Message Passing Interface (MPI) ranks to achieve parallel simulation.
Field programmable gate array (FPGA)-based emulators run significantly faster
than simulation software but require a huge amount of effort to develop and
validate register-transfer level models \cite{ISCA18:FireSim}.
In comparison, our work accelerates simulation from a different angle to make
the most of widely available ML accelerators, such as GPUs.

\secspace
\section{Conclusions}
\label{sect:conclusion}
\secspace

%
%

This work proposes a new computer architecture simulation paradigm using ML.
To the best of our knowledge, this effort is the first to demonstrate ML's
applicability to full-fledged architecture simulation.
This new methodology significantly improves simulation performance without
sacrificing accuracy.
In addition to discrete-event, analytical, or other statistical approaches to
architectural simulation/modeling, we maintain this new class of simulators will
become a useful, valuable addition to the architect's ``bag-of-tools.''
We recognize several advantages of this new approach.

1) We demonstrate that ML-based simulators can predict overall performance
accurately, and they also qualitatively capture architecture and application
behaviors.
2) ML is intrinsically easier to parallelize than discrete-event simulation.
Moreover, ML-based simulators capitalize on modern computing technology that is
tailored for boosting ML performance.
3) ML-based simulators generalize well to a large spectrum of application workloads.
In our approach, this stems from building them around an instruction-level
latency predictor. Hence, the focus is on learning instruction behaviors rather
than high-level program behaviors that are much more difficult to capture.
4) The training data are easy and fast to obtain. Potential sources of training
data are multiple, including simplified models of simulators, actual execution
of code on existing systems, or historical performance data.

\subhead{Future Directions.}
We plan to investigate ML-based approaches that support multi-thread/program
simulation as our next step.
The key to supporting multi-thread/program simulation is to model
communications.
We describe two possible strategies as follows, 1) extending context
instructions to include concurrently executed instructions from other
threads/programs, and 2) training ML models to model the impact of shared
resources (e.g., caches, memory).

\begin{acks}
We would like to thank the anonymous reviewers for their helpful comments and
Sergey Blagodurov for shepherding this paper.
This research was conducted at the Brookhaven National Laboratory, supported by
the U.S. Department of Energy’s Office of Science under Contract No.
DE-SC0012704.
\end{acks}

\bibliographystyle{ACM-Reference-Format}
\bibliography{ref}


\end{document}